\documentclass[usenatbib]{mn2e}
\usepackage{amsmath} 
\usepackage{epsfig}

\newcommand{\cy}[1]{\citeyear{#1}}  

\title[The  X-ray  fast-time variability  of  Sco~X-2]{The X-ray  fast-time
variability of Sco~X-2 (GX~349+2) with \textbfit{RXTE\/}}

\author[P.    M.    O'Neill   et  al.]    {P.~M.~O'Neill,$^1$\thanks{email:
p.m.oneill@adfa.edu.au}     E.~Kuulkers,$^{2,3}$     R.~K.~Sood$^1$     and
M.~van~der~Klis$^4$\\  $^1$School  of  Physics,  Australian  Defence  Force
Academy, Canberra  ACT 2600,  Australia \\ $^2$Space  Research Organization
Netherlands,  Sorbonnelaan   2,  3584   CA  Utrecht,  The   Netherlands  \\
$^3$Astronomical Institute,  Utrecht University,  P.O.  Box 80000,  3507 TA
Utrecht, The  Netherlands \\ $^4$Astronomical  Institute `Anton Pannekoek',
University of Amsterdam, Kruislaan  403, 1098 SJ Amsterdam, The Netherlands
}

\date{Accepted. Received}

\pagerange{\pageref{firstpage}--\pageref{lastpage}} \pubyear{2002}

\begin{document}

\maketitle

\label{firstpage}

\begin{abstract}

Sco~X-2  (GX~349+2) is  a  low-mass X-ray  binary  and Z  source.  We  have
analysed $\sim$156~ks of {\it Rossi X-ray Timing Explorer\/} data, obtained
in 1998 January, on this source.  We investigated the fast-time variability
as  a function  of position  on the  Z track.   During  these observations,
Sco~X-2  traced out  the most  extensive Z  track ever  reported  from this
object, making  this the most comprehensive  study thus far.   We found the
broad  peaked flaring  branch  noise that  is  typical of  Sco~X-2, with  a
centroid  frequency  in the  range  3.3--5.8~Hz.   We  also discovered  low
frequency noise, and a new  peaked noise feature, with centroid frequencies
in  the range  5.4--7.6~Hz  and 11--54~Hz,  respectively.   We discuss  the
phenomenology of these features, their relationship with the power spectral
components found in other low-mass X-ray binaries, and the implications for
current models.   In particular,  the low frequency  noise we  observed was
strongest at intermediate energies, in  contrast to the low frequency noise
seen in  other Z sources.  We  also detected very low  frequency noise, and
have calculated  complex cross spectra between intensity  and hardness.  We
found that the very low frequency noise is not entirely due to motion along
the Z track.

\end{abstract}


\section{Introduction \label{sect:intro}}

Sco~X-2  is one  of the  persistently  bright neutron  star low-mass  X-ray
binaries known  as Z sources.  Z  sources are so-named  because, over time,
they  trace out  a `Z'  shaped  track in  an X-ray  colour-colour (CD)  and
hardness-intensity  (HID)  diagram.  The  top  of  the  `Z' is  called  the
horizontal branch  (HB), the  diagonal is the  normal branch (NB),  and the
bottom  is the flaring  branch (FB)  (Hasinger \&  van~der~Klis \cy{hv89}).
Inferred mass-accretion rate increases from  the HB, through the NB, to the
FB.  The mass-accretion rate is thought to reach the Eddington limit at the
bottom  of the  NB,  and is  super-Eddington  in the  FB.   The five  other
traditional  Z  sources  are:   Sco~X-1,  GX~17+2,  Cyg~X-2,  GX~340+0  and
GX~5$-$1.  A related class of objects are the atoll sources; they exhibit a
curved  track   in  a  colour-colour  diagram   (Hasinger  \&  van~der~Klis
\cy{hv89}).

\subsection{Z and atoll source fast-time variability \label{sec:zaasfv}}

Through  the study  of the  X-ray fast-time  variability of  low-mass X-ray
binaries,  we   are  able  to  investigate   mass-accretion  processes;  in
particular, the fast-time variability is  a signature of the flow of matter
near the  compact object (see  review van~der~Klis \cy{v00}).   The typical
power  spectral properties  of  Z  and atoll  sources  are correlated  with
position in  the CD (Hasinger \& van~der~Klis  \cy{hv89}; also van~der~Klis
\cy{v95a}).

On the HB, Z sources  typically exhibit six variability components that are
phenomenologically  related (e.g.,  Homan~et~al. \cy{hvj02}  and references
therein):  a band-limited  noise component  with cut-off  frequency  in the
range  2--10~Hz,  called  low   frequency  noise  (LFN);  a  quasi-periodic
oscillation (QPO)  in the range  15--60~Hz, known as the  horizontal branch
oscillation (HBO);  a broad sub-HBO peak  located between the  LFN and HBO,
and  which has  been identified  as  a sub-harmonic  of the  HBO; a  second
harmonic to the  HBO; and a pair of QPOs in  the range 200--1130~Hz, called
kHz QPOs.

On the  NB and lower part  of the FB there  is a QPO in  the range 6--20~Hz
(e.g.,  van~der~Klis \cy{v95a};  Dieters \&  van~der~Klis  \cy{dv00}; Homan
et~al. \cy{hvj02}).  On the NB, it  is known as a normal branch oscillation
(NBO),  and its  frequency is  roughly constant  near 6~Hz.   With movement
through the NB/FB vertex  the mean frequency and full-width-at-half-maximum
(FWHM)  of the QPO  rise rapidly.   It is  then known  as a  flaring branch
oscillation  (FBO),  and  becomes  indistinguishable  from  the  underlying
continuum at about 10--20~per~cent of the way up that branch.

Some of the features  seen in the HB are also sometimes  seen on the NB and
FB.   In  Sco~X-1, van~der~Klis~et~al.   (\cy{vwh97})  found  that the  NBO
appeared  to   `peak-up'  out  of   the  LFN.   In   GX~17+2,  Homan~et~al.
(\cy{hvj02}) found  that the NBO  and LFN existed simultaneously,  and were
separate features.  They also observed a peaked noise feature in the FB, in
addition to an FBO, which they identified as LFN.

A power-law noise component, called very low frequency noise (VLFN), exists
throughout the Z track, and  is stronger at higher mass-accretion rates and
at higher energies (e.g., van~der~Klis \cy{v95a}).  van~der~Klis (\cy{v91})
suggested  that it  is produced  by the  variations in  intensity  that are
associated  with  movement along  the  Z  track.   Dieters \&  van~der~Klis
(\cy{dv00})    constructed   power    spectra   of    rank    number   (see
Section~\ref{sect:cohidaps}),       in       the      frequency       range
3$\times$10$^{-5}$--0.01~Hz, and  also of intensity, in  the range 8$\times
10^{-4}$--0.2~Hz.   They found  that the  power-law indices  from  the rank
number power  spectra were consistent  with those from the  intensity power
specta,  although  we  note  that  the uncertainties  on  the  rank  number
power-law indices were quite large.  If VLFN is due to movement along the Z
track, then the  speed of that movement should be  correlated with the VLFN
strength.  Dieters \& van~der~Klis (\cy{dv00}) found this to be the case in
Sco~X-1, Homan  et~al.  (\cy{hvj02}) found  that the correlation  exists in
some  parts of the  Z track  of GX~17+2,  and Wijnands  et~al. (\cy{wvk97})
found that  the VLFN  strength in Cyg~X-2  did not vary  monotonically with
speed.

Finally,  a  noise  component  with   a  cut-off  frequency  in  the  range
30--100~Hz,  called high  frequency  noise (HFN),  has  also been  observed
(e.g., van~der~Klis \cy{v95a}).

At  their lowest  inferred mass-accretion  rates atoll  sources  exhibit: a
band-limited  noise component,  modelled using  a broken  power-law  with a
break frequency  in the range 0.1--32~Hz, called  (rather confusingly) high
frequency noise; low  frequency QPOs or broad peaks  in the range 1--67~Hz;
and kHz QPOs.  Except for the  kHz QPOs, these features are very similar to
those seen in black hole  candidates (BHCs) (e.g., Wijnands \& van~der~Klis
\cy{wv99}  and references  therein).   In atoll  souces  VLFN dominates  at
higher mass-accretion rates.

A  correlation  was  found in  Z  sources  and  atoll sources  between  the
frequencies  of  the kHz  QPOs,  and  the HBO  or  1--67~Hz  QPO; the  same
correlation  was  found between  features  in  the  power spectra  of  BHCs
(Psaltis,   Belloni  \&  van~der~Klis   \cy{pbv99};  also   Psaltis  et~al.
\cy{pwh99}).   Wijnands  \&  van~der~Klis  (\cy{wv99})  discovered  another
correlation in atoll  sources and BHCs, between the  break frequency of the
band-limited noise (the  HFN in atoll sources), and  the centroid frequency
of  a QPO  or  bump in  the range  0.2--67~Hz  (the 1--67~Hz  QPO in  atoll
sources).  The same correlation was found to exist in Z sources between the
LFN and  sub-HBO peak, and  also between the  sub-HBO and HBO  (Wijnands \&
van~der~Klis  \cy{wv99};  also  Homan  et~al.   \cy{hvj02}).   van~der~Klis
(\cy{v94a},b), Belloni,  Psaltis \& van~der~Klis  (\cy{bpv02}), and Sunyaev
\&  Revnivtsev (\cy{sr00})  have also  made comparisons  between  the power
spectra of BHCs and neutron star LMXBs.

There is  currently much effort being  put into developing  models that can
explain  these  correlations.   The   discovery  that  the  power  spectral
components in black hole candidates  follow the same correlations seen in Z
sources and atoll sources gives support to those models which do not depend
on the  presence or otherwise of  a solid surface or  magnetic field (e.g.,
van~der~Klis \cy{v00}).

\subsection{Scorpius~X-2 (GX~349+2) \label{sec:sx}}

Except for  the fact that Sco  X-2 has never  been observed in the  HB, its
hardness-intensity diagram looks nearly identical to that of Sco~X-1 (e.g.,
Schulz, Hasinger  \& Tr\"{u}mper \cy{sht89}).  Therefore,  one might expect
the fast-time variability of Sco~X-2 to be similar to that seen in Sco~X-1.

Ponman, Cooke \& Stella (\cy{pcs88}) analysed {\it EXOSAT\/} data and found
broad  peaked  noise  in  the  FB,  with a  centroid  frequency,  FWHM  and
fractional   rms  amplitude   of  roughly   5~Hz,  10~Hz   and  3~per~cent,
respectively.  Kuulkers  \& van~der~Klis (\cy{kv98}) analysed  a small {\it
RXTE\/} dataset and  found that the peaked noise  became narrower and moved
to lower frequencies  with movement from the NB into  the FB.  O'Neill et~al.
(\cy{oks01}) analysed  {\it Ginga\/}  data, and found  peaked noise  with a
centroid   frequency  and  FWHM   in  the   range  4--7~Hz   and  6--12~Hz,
respectively.  They found that it was strongest at about 10~per~cent of the
way up FB,  and was present until halfway up the  branch.  They referred to
the feature as  flaring branch noise (FBN), and concluded  that: FBN is not
the same as  atoll source high frequency noise; it  is difficult to explain
it with the model  for typical normal/flaring branch oscillations (Fortner,
Lamb \& Miller  \cy{flm89}); and it resembles the peaked  noise seen in the
FB of Cyg~X-2 at low overall intensities.  O'Neill et~al. (\cy{oks01}) also
found very  low frequency noise that  became stronger with  movement up the
FB, as seen in other Z sources.

Zhang, Strohmayer \& Swank (\cy{zss98}) observed Sco~X-2 with {\it RXTE} in
1998  January.   During those  observations  Sco~X-2  traced  out the  most
extensive  Z track ever  seen from  that object;  most importantly,  it was
found  higher  up  in  the  NB  than ever  before.   A  pair  of  kilohertz
quasi-periodic   oscillations   were  detected   at   712~Hz  and   978~Hz,
respectively,  in  the uppermost  part  of  the  observed NB.   Agrawal  \&
Bhattacharyya (\cy{ab01}) analysed $\sim$40~ks  of {\it RXTE} data obtained
in 1998 September and October and also found an extended NB.  They detected
peaked noise  similar to  that found in  previous observations,  though its
fractional  rms  amplitude did  not  vary  smoothly,  and the  feature  was
detected as high as 80~per~cent of the  way up the FB.  They found that the
properties of  the peaked  noise were similar  throughout the Z  track, and
thus concluded that  the peaked noise seen  in the NB was the  same as that
seen in the  FB.  A QPO, with a centroid frequency  of 3.8~Hz, was detected
(90~per~cent confidence level) in the FB during a dip in the intensity that
occurred after a flare.  During  the September and October observations the
behaviour of the  VLFN was very different to that  seen previously: it was,
on average,  stronger in  the NB  than in the  FB, and  did not  exhibit an
increase in strength with movement up the FB.

On the basis  of phenomenology, the peaked noise observed in  the NB and FB
of  Sco~X-2 is  unlikely to  be an  N/FBO (e.g.,  Agrawal  \& Bhattacharyya
\cy{ab01} and references therein).  First,  the peaked noise seen in the NB
of Sco~X-2 is much broader than  an NBO.  Second, as Sco~X-2 moves from the
NB, through  the NB/FB vertex, and  then up the FB,  the centroid frequency
and  FWHM of the  peaked noise  decrease, and  then stay  roughly constant.
This is in contrast to an N/FBO, of which, with the same movement along the
track, the centroid frequency and FWHM increase rapidly.  Third, the peaked
noise in Sco~X-2 has been seen as  far as 80~per~cent of the way up the FB,
while FBOs are only present in the lowest 10--20~per~cent of that branch.

Zhang  et~al. (\cy{zss98})  did  not analyse  the  low frequency  fast-time
variability  in detail.  We,  therefore, decided  to reanalyse  those data.
Our primary objectives were: to look  for the LFN and HBO that are expected
to accompany  kHz QPOs; to better  understand the phenomenology  of FBN; to
study the properties of the kHz QPOs  as a function of both position on the
Z track and photon energy; and, to further investigate VLFN.  A preliminary
account of this work has been given by O'Neill et~al. (\cy{oksb}).

\section{Observations and Analysis \label{sect:oban}}

The {\it Rossi X-ray Timing Explorer\/} ({\it RXTE\/}; Bradt, Rothschild \&
Swank \cy{brs93}) proportional counter array (PCA) observed Sco~X-2 between
1  and   29  January  1998.    The  $\sim$156~ks  of  data   were  recorded
simultaneously in  six different modes.  The  `Standard 2' mode  data had a
time  resolution of  16~s,  and Sco~X-2  was  visible in  the energy  range
2.0--40~keV.  Three `single bit' modes and one `event mode' were used, with
a time  resolution of 122~$\mu$s.   The effective energy boundaries  of the
single bit and event mode data were 2.0--5.1--6.5--8.7~keV and 8.7--40~keV,
respectively.  We  only used  data in which  all five  proportional counter
units were operating.

\subsection{Construction of hardness-intensity diagram and power spectra
\label{sect:cohidaps}}

\begin{figure}
\begin{center}
\epsfig{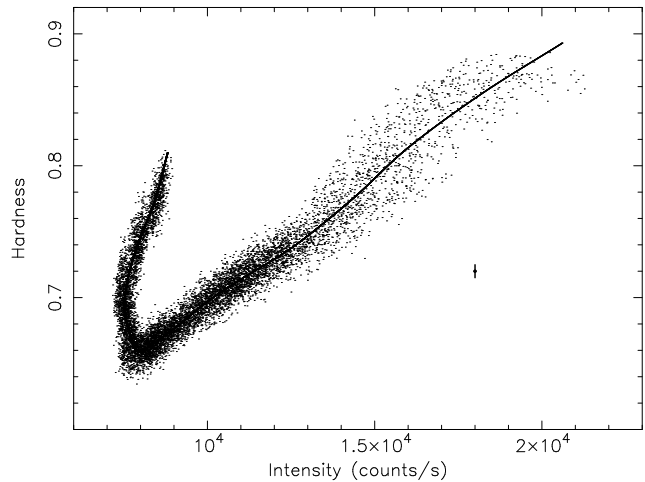}
\caption{Hardness-intensity diagram of Standard 2 data (16~s averages). 
The  hardness is  the counting  rate  ratio between  the 8.7--19.7~keV  and
6.2--8.7~keV bands,  and the  intensity is the  counting rate in  the range
2.0--19.7~keV. The solid  line shows the spline used  to calculate S$_{z}$.
A representative error bar is shown.}
\label{fig:b_rk}
\end{center}
\end{figure}

A hardness-intensity diagram was constructed from the Standard 2 data using
16~s  averages, and is  shown in  Fig.~\ref{fig:b_rk}.  The  counting rates
were  corrected for  detector deadtime  and  background.  In  the HID,  the
hardness is  defined as the counting  rate ratio between  the energy ranges
8.7--19.7~keV and 6.2--8.7~keV,  and the intensity is the  counting rate in
the  range 2.0--19.7~keV.   Fig.~\ref{fig:b_rk} shows  the most  complete Z
track ever observed  from Sco~X-2 (see also Zhang  et~al. \cy{zss98}).  The
counting rate  at the top  of the FB  is a factor  of 2.6 greater  than the
intensity at the NB/FB vertex.  In  comparison, the factor was: 2.2 in {\it
Ginga \/} data (1.1--16.7~keV,  64~s averages; O'Neill et~al.  \cy{oks01});
$\sim$2.1 in  the {\it  EXOSAT \/} data  (1--10~keV, 64~s  averages; Ponman
et~al.  \cy{pcs88});  and $\sim$2  in the 1998  September and  October {\it
RXTE  \/}  data  (2--16~keV,   256~s  averages;  Agrawal  \&  Bhattacharyya
\cy{ab01}).   We note  that using  lower energy  bands and  longer averages
reduces the observed maximum intensity of the FB.

Light curves were extracted from the single bit and event mode data, with a
time  resolution of  244~$\mu$s.  We  summed  the energy  resolved data  to
produce  light curves  with  various energy  ranges.   We calculated  power
spectra from 16~s intervals in each light curve; each interval corresponded
exactly  to  one point  in  the  HID.   To investigate  the  0.0625--300~Hz
frequency range we used a  logarithmic frequency rebinning scheme, while we
used linear rebinning to investigate the kilohertz region.

A spline was  fitted through the HID, and rank  numbers were calculated for
each point.  Rank number is a  one dimensional measure of position in the Z
track; it was introduced by Hasinger et~al.  (\cy{hve90}), and is symbolised
as  S$_{z}$ (Hertz  et~al.  \cy{hvw92};  see also  Dieters  \& van~der~Klis
\cy{dv00}).   It is  a measure  of the  inferred mass-accretion  rate.  We
defined the NB/FB vertex as S$_{z}$=2, and the top of the FB as S$_{z}$=3.

Average power  spectra were then constructed  on the basis  of rank number,
and were normalised  to give fractional rms amplitude  squared per Hz.  The
error given  for each mean  S$_{z}$ is the  standard deviation of  the rank
numbers used in the average.

\subsection{Fitting of power spectra \label{sect:fops}}

To  investigate the  low  frequency fast-time  variability,  we fitted  the
resultant power spectra  in the range 0.0625--300~Hz using:  a power-law to
represent very  low frequency noise,  Lorentzians as required  to represent
peaked noise, and a constant to represent the deadtime-affected white-noise
level.   We  investigated  the  kilohertz  quasi-periodic  oscillations  by
fitting the  resultant power spectra  in the range 500--1200~Hz,  using two
Lorentzians and a constant.

The  VLFN  fractional  rms  amplitude  was measured  by  extrapolating  the
best-fitting  power-law and  integrating  over the  range 0.01--1~Hz.   The
addition of  a Lorentzian  component to a  model was deemed  necessary only
when the probability of  exceeding the $\chi_{\nu}^{2}$ of the best-fitting
model was $<$0.001, and an F-test for the inclusion of the extra Lorentzian
was significant  at or  above the 3$\sigma$  level.  We fixed  the centroid
frequency and/or the FWHM of  a Lorentzian component either when they could
not  be constrained, or  when a  fractional rms  amplitude upper  limit was
being measured.  Uncertainties in the best-fitting parameters were measured
using   $\Delta\chi^{2}=1$,   and   upper   limits  were   measured   using
$\Delta\chi^{2}=2.71$ (95~per~cent confidence level).

We corrected  the best-fitting  fractional rms amplitudes  for differential
deadtime and channel cross-talk  as appropriate (see Appendix~A).  Finally,
we corrected the fractional rms amplitudes for the background counting rate
(e.g., van~der~Klis  \cy{v95b}).  In the  low frequency power  spectra, the
reduction of the rms due to the binning of data was negligible; the binning
correction factor applied to the rms  of the kHz QPOs was 1.1 (van~der~Klis
\cy{v89}).

\subsection{Investigation of very low frequency noise \label{sect:iovlfn}}

To further investigate  the VLFN, we constructed a HID  from the single bit
and event  mode data using 32~s  averages.  We defined the  hardness as the
counting rate ratio between the energy ranges 8.7--40~keV and 6.5--8.7~keV,
and  the intensity  as  the counting  rate  in the  range 2.0--40~keV.   We
calculated complex Fourier  spectra from both the hardness  light curve and
the deadtime corrected intensity light curve, using 1~s time resolution and
32~s intervals.  Each interval, along with its associated Fourier spectrum,
corresponded exactly to one point in  the HID.  We used 1~s time resolution
so  we  could  directly   investigate  the  fast-time  variability  in  the
time-domain.  The Fourier spectra had a frequency range of 0.03125--0.5~Hz.

For  each 32~s  interval we  calculated complex  cross spectra  between the
hardness and  intensity.  We multiplied  the complex Fourier  amplitudes of
the intensity, by  the complex conjugate of the  complex Fourier amplitudes
of the hardness (e.g.,  van~der~Klis et~al. \cy{vhs87}).  At each frequency
in  the resultant  cross spectrum  there  is a  corresponding cross  vector
$\bmath{C}$, with a  real and imaginary component.  The  phase of the cross
vector is the  phase difference between the variations  in hardness and the
variations  in  intensity;  a  negative  phase  means  that  hardness  lags
intensity.  A phase of 0$\degr$ means hardness and intensity are positively
correlated, and a phase of 180$\degr$ means they are anti-correlated. 

We then constructed  average complex cross spectra.  At  each frequency, we
averaged together the real and  imaginary components from all cross spectra
whose  corresponding point  in  the HID  was  within a  specified range  of
hardness and intensity.  The mean amplitude and phase at each frequency was
calculated from the mean  real and imaginary components.  The uncertainties
in  the  mean real  and  imaginary  components  are the  observed  standard
deviation  in  the  mean  of   those  components,  and  we  determined  the
uncertainties  in amplitude  and  phase by  using  error propagation.   The
phases at the Nyquist frequency are always either 0$\degr$ or 180$\degr$ so
they have no error bars.

If, at a particular frequency, the hardness and intensity are uncorrelated,
then the  phases of the  {\it individual\/} cross vectors  corresponding to
that frequency will be uniformly  distributed between $-\pi$ and $+\pi$. In
this  situation, the  real and  imaginary components  of  the corresponding
vector  $\bmath{G}$ in  the  {\it  average\/} cross  spectrum  will have  
Gaussian  distributions  with  a  mean  of  zero.   The  real  component  of
$\bmath{G}$ has a variance, $\sigma_{\mathrm{R}}^{2}$, that is equal to the
variance of the imaginary component, $\sigma_{\mathrm{I}}^{2}$.  These may
be calculated by using the expression

\begin{equation}
\sigma_{\mathrm{R}}^{2}       =       \frac{\sigma_{|\bmath{C}|}^{2}      +
\mu_{|\bmath{C}|}^{2}}{2M}
\end{equation}

\noindent where $\sigma_{|\bmath{C}|}$ and $\mu_{|\bmath{C}|}$ are the
observed standard  deviation and mean,  respectively, of the  amplitudes of
the set of cross vectors that were averaged together, and $M$ is the number
of vectors comprising that average.

Therefore, in the absence of  a correlation between intensity and hardness,
the  squared amplitude  $|\bmath{G}|^{2}$ of  the average  cross  vector is
distributed  such that  the  probability of  observing a  $|\bmath{G}|^{2}$
greater  than a certain  threshold level  $T$ is  given by  (Dieters et~al.
\cy{dvk00}; Dieters, private communication)

\begin{equation}
\mathrm{Prob}(|\bmath{G}|^{2}              >              T)              =
\mathrm{exp}(-T/|\bmath{G_{\mathrm{0}}}|^{2})
\end{equation}

\noindent where 

\begin{align}
|\bmath{G_{\mathrm{0}}}|^{2}       &=       \sigma_{\mathrm{R}}^{2}       +
 \sigma_{\mathrm{I}}^{2}\\ 
 &   =  2   \sigma_{\mathrm{R}}^{2}\\   |\bmath{G_{\mathrm{0}}}|^{2}  &   =
 \frac{\sigma_{|\bmath{C}|}^{2} + \mu_{|\bmath{C}|}^{2}}{M} 
\end{align}

\noindent Given an observed squared amplitude $T$, the significance of the
 detection of a correlation, expressed as a percentage, is given by

\begin{equation}
\mathrm{significance}        =         100        \times        (1        -
\mathrm{exp}(-T/|\bmath{G_{\mathrm{0}}}|^{2}))
\end{equation}

The cross vector at the Nyquist  frequency has only a real component, so it
must         be        treated        separately.          The        value
$|\bmath{G}|^{2}/\sigma_{\mathrm{R}}^{2}$ is distributed as $\chi^{2}$ with
1     degree    of     freedom,     where    $\sigma_{\mathrm{R}}^{2}     =
(\sigma_{|\bmath{C}|}^{2} + \mu_{|\bmath{C}|}^{2}) / M$.

For a comparison with the analytical significance levels, we also used a
Monte  Carlo  simulation to  determine  the  significance  of the  observed
correlations.  We randomised the  phase difference between the hardness and
intensity  at each  frequency, in  each {\it  individual \/}  complex cross
spectra,  while maintaining the  observed amplitudes.   We then  used those
`random-phase' cross  spectra to construct  an average cross  spectrum.  We
repeated this  procedure 10000 times  and obtained, at each  frequency, the
distribution  of mean amplitudes  that would  be present  if our  data were
actually uncorrelated.  With this method, the significance of a correlation
is equal  to the  percentage of random-phase amplitudes that  are less
than the observed amplitude.

We  deemed the hardness  and intensity  to be  correlated, at  a particular
frequency, if the significance obtained from the analytical method was at
least 90~per~cent.   The phases of  those cross vectors with  a significant
correlation  are  presented   in  Table~\ref{tab:phases},  along  with  the
significance we obtained  from each method.  The results  obtained from the
Monte Carlo method were, on average, within 0.3 per~cent significance of
those found from the analytical method (see Table~\ref{tab:phases}). 

We repeated the entire procedure  twice.  First, using a time resolution of
8~s   with  256~s   intervals,   corresponding  to   the  frequency   range
0.00390625--0.0625~Hz, which allowed us  to investigate the frequency range
in  which  the  power-law   component  dominates.   Second,  using  a  time
resolution of  0.125~s with 4~s  intervals, corresponding to  the frequency
range 0.25--4.00~Hz, which  allowed us to investigate the  influence of low
frequency noise and flaring branch noise on our cross spectra.

\section{Fast-time variability}

\begin{figure}
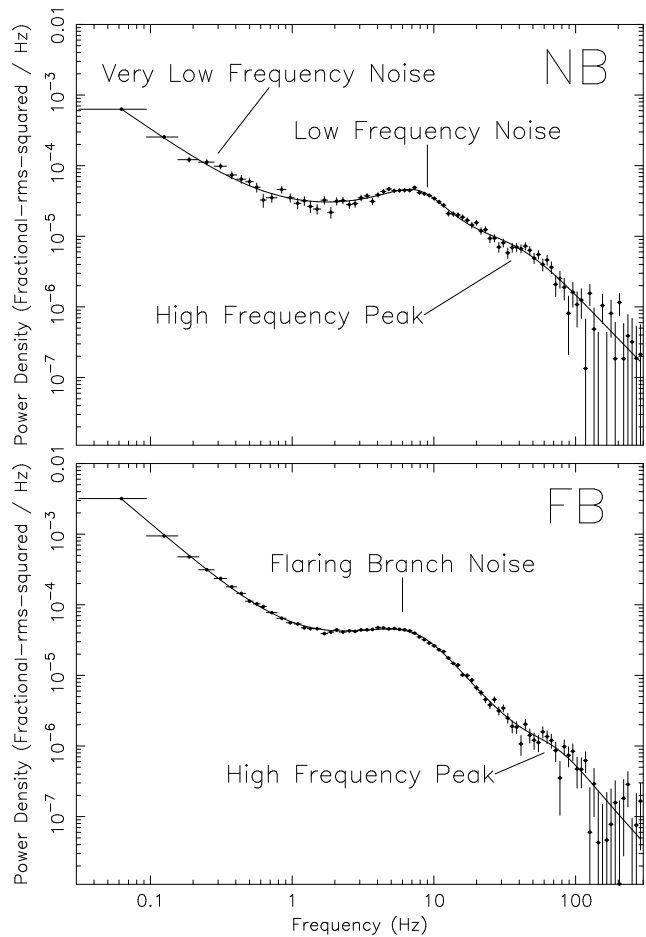

\begin{center}
\epsfig{file=MC390_fig2_top.ps,height=84mm,angle=270}
\epsfig{file=MC390_fig2_bottom.ps,height=84mm,angle=270}
\caption{Representative power spectra (2.0--40~keV) from the NB
  (S$_{z}$  range  1.53--1.85)  and  FB  (S$_{z}$  range  2.03--3.0).   The
white-noise  levels have  been  subtracted  and the  solid  lines show  the
best-fitting models.}
\label{fig:f_55_s}
\end{center}
\end{figure}

In  Fig.~\ref{fig:f_55_s} we  present  representative 0.0625--300~Hz  power
spectra from  the NB (Fig.~\ref{fig:f_55_s}~top;  S$_{z}$ range 1.53--1.85)
and  FB (Fig.~\ref{fig:f_55_s}~bottom;  S$_{z}$ range  2.03--3.0),  and the
various  components we identified.   We detected  three features:  very low
frequency  noise; low  frequency  noise/flaring branch  noise;  and a  high
frequency peaked noise component.

We  subdivided our  data  further and  constructed  thirteen average  power
spectra  covering the  entire Z  track.  The  best-fitting values  from the
thirteen average power spectra,  in the range 0.0625--300~Hz, are presented
in Table~\ref{tab:powfits2}.

In  Table~\ref{tab:powfits2} we  also present  the  $\nu_{\mathrm{max}}$ of
each Lorentzian  component.  This can  be calculated for both  a Lorentzian
and  cut-off power-law,  and it  facilitates a  comparison between  the two
forms.   The $\nu_{\mathrm{max}}$  is the  frequency  at which  there is  a
maximum in a power-density-times-frequency plot, and it is a useful measure
of the characteristic frequency of  a power spectral feature (e.g., Belloni
et~al. \cy{bpv02}).

The   kilohertz   quasi-periodic  oscillations   were   only  detected   at
S$_{z}$=1.66.   In  Table~\ref{tab:khzfits}  we  present  the  best-fitting
values of the kHz QPOs in the NB.  We investigated the energy dependence of
the  kHz QPOs  by fitting  energy resolved  power spectra  corresponding to
S$_{z}$=1.66.  We  fixed the  centroid frequencies and  FWHM at  the values
found from the 2.0--40~keV power spectrum.  In the energy resolved spectra,
we only  detected kHz  QPOs in the  5.1--40~keV range, with  fractional rms
amplitudes of  2.3$\pm$0.2~per~cent and 2.1$\pm$0.2~per~cent  for the lower
and upper peaks, respectively.  In the energy range 2.0--8.7~keV, the upper
limits  of the  rms for  the lower  and upper  peaks were  2.1~per~cent and
1.8~per~cent, respectively.   Therefore, the kHz QPOs in  Sco~X-2 are hard,
like  those  seen  in   other  objects  (e.g.,  van~der~Klis  \cy{v00}  and
references therein).  Zhang et~al.   (\cy{zss98}) fitted the kHz QPOs using
Gaussians; we also  fitted our QPOs using Gaussians  and found best-fitting
values  consistent  with their  results.   The $\chi_{\nu}^{2}$/degrees  of
freedom (dof) was 0.97/51 for the Gaussian model, compared with 0.93/51 for
the Lorentzian model.

To investigate  the energy dependence of the  0.0625--300~Hz power spectral
features we used energy resolved power spectra corresponding to the average
spectra shown in Fig.~\ref{fig:f_55_s}.   We fixed the centroid frequencies
and FWHM at the values found from the 2.0--40~keV spectra.  The results are
shown in Table~\ref{tab:enefits}.

In the next three subsections we  describe in more detail the properties of
the LFN/FBN peak, the high frequency peak, and the VLFN.

\begin{table*}
\begin{flushleft}
\caption{Best-fitting values from average power spectra (2.0--40~keV).} 
\label{tab:powfits2}
\setlength{\tabcolsep}{4pt}
\begin{tabular}{cccccccccccc} \hline 
 S$_{z}$    &    \multicolumn{2}{c}{VLFN}    &
 \multicolumn{4}{c}{LFN~/~FBN}   &
 \multicolumn{4}{c}{High Frequency Peak}  & $\chi^{2}_{\nu}$/dof \\ & Index
 &  rms &  Freq  &  FWHM &  $\nu_{\mathrm{max}}$  & rms  &  Freq  & FWHM  &
 $\nu_{\mathrm{max}}$  & rms  \\ &  & (per~cent)  & (Hz)  & (Hz)  &  (Hz) &
 (per~cent) & (Hz) & (Hz) & (Hz) & (per~cent) \\ \hline

1.66$\pm$0.04 & 1.55$\pm$0.07 &  1.40$\pm$0.05 & 5.4$\pm$0.2 & 10.3$\pm$1.5
  &  7.5$\pm$0.5  &  2.5$\pm$0.2  &  22$\pm$11 &  70$\pm$11  &  41$\pm$7  &
  3.0$\pm$0.5 & 1.37 / 76 \\

1.79$\pm$0.04 & 1.37$\pm$0.06 & 1.33$\pm$0.04 & 7.6$\pm$0.2 & 7.3$\pm$0.8 &
8.4$\pm$0.2 & 2.3$\pm$0.2 & 11$\pm$5 & 57$\pm$11 & 30$\pm$5 & 3.3$\pm$0.3 &
1.25 / 76 \\

1.91$\pm$0.04 & 1.24$\pm$0.04 & 1.32$\pm$0.03 & 7.6$\pm$0.2 & 8.0$\pm$1.3 &
8.6$\pm$0.4 & 2.0$\pm$0.2 & 14$\pm$4  & 32$\pm$6 & 21$\pm$3 & 2.3$\pm$0.3 &
1.09 / 76 \\

2.00$\pm$0.02 & 1.56$\pm$0.04 &  1.83$\pm$0.05 & 5.8$\pm$0.1 & 11.8$\pm$0.3
  & 8.3$\pm$0.1 & 4.06$\pm$0.04 & 14$^a$  & 32$^a$ & 21$^a$ & $<$1.0 & 0.98
  / 79 \\

2.05$\pm$0.01 & 1.68$\pm$0.06 &  1.99$\pm$0.08 & 5.5$\pm$0.1 & 11.7$\pm$0.3
  & 8.0$\pm$0.1 & 4.33$\pm$0.05 & 14$^a$  & 32$^a$ & 21$^a$ & $<$1.7 & 1.14
  / 79 \\

2.09$\pm$0.01 & 1.76$\pm$0.06 &  2.20$\pm$0.09 & 5.2$\pm$0.1 & 11.6$\pm$0.3
  & 7.8$\pm$0.1 & 4.13$\pm$0.05 & 14$^a$  & 32$^a$ & 21$^a$ & $<$1.4 & 1.07
  / 79 \\

2.15$\pm$0.02 & 1.74$\pm$0.05 &  2.35$\pm$0.08 & 5.4$\pm$0.1 & 10.3$\pm$0.3
  & 7.5$\pm$0.1 & 3.72$\pm$0.04 & 14$^a$  & 32$^a$ & 21$^a$ & $<$1.6 & 1.03
  / 79 \\

2.21$\pm$0.01 & 1.68$\pm$0.04 &  2.41$\pm$0.08 & 5.5$\pm$0.1 & 11.6$\pm$0.4
  & 8.0$\pm$0.2 & 3.30$\pm$0.05 & 14$^a$  & 32$^a$ & 21$^a$ & $<$1.7 & 1.52
  / 79 \\

2.26$\pm$0.02 & 1.72$\pm$0.04 &  2.78$\pm$0.09 & 4.6$\pm$0.3 & 14.3$\pm$0.7
  & 8.5$\pm$0.3 & 3.01$\pm$0.07 & 14$^a$  & 32$^a$ & 21$^a$ & $<$1.7 & 0.82
  / 79 \\

2.32$\pm$0.02 &  1.71$\pm$0.04 & 2.99$\pm$0.09  & 4.4$\pm$0.6 &  16$\pm$1 &
9.1$\pm$0.5 & 2.52$\pm$0.09 & 14$^a$ & 32$^a$ & 21$^a$ & $<$1.4 & 1.34 / 79
\\

2.41$\pm$0.03 &  1.73$\pm$0.04 & 3.23$\pm$0.12  & 3.3$\pm$1.5 &  16$\pm$3 &
8.7$\pm$1.5 & 1.9$\pm$0.2 &  14$^a$ & 32$^a$ & 21$^a$ & $<$1.3  & 0.96 / 79
\\

2.55$\pm$0.05  &  1.64$\pm$0.03 &  3.09$\pm$0.10  &  3.3$^a$  & 11$\pm$2  &
6.4$\pm$0.9 & 1.3$\pm$0.1 &  14$^a$ & 32$^a$ & 21$^a$ & $<$1.1  & 1.14 / 80
\\

2.77$\pm$0.09 & 1.60$\pm$0.02 & 5.45$\pm$0.12  & 3.3$^a$ & 11$^a$ & 6.4$^a$
& $<$0.8 & 14$^a$ & 32$^a$ & 21$^a$ & $<$0.8 & 0.97 / 82 \\

\hline
\end{tabular}
\\
\noindent $^a$Fixed parameter.
\end{flushleft}
\end{table*}

\begin{table*}
\begin{flushleft}
\caption{Best-fitting kilohertz QPO values from NB average power
  spectra (2.0--40~keV).}
\label{tab:khzfits}
\setlength{\tabcolsep}{4pt}
\begin{tabular}{cccccccc} \hline 
 S$_{z}$ & \multicolumn{3}{c}{Lower kHz QPO} & \multicolumn{3}{c}{Upper kHz
 QPO} & $\chi^{2}_{\nu}$/dof \\ & Freq  (Hz) & FWHM (Hz) & rms (per~cent) &
 Freq (Hz) & FWHM (Hz) & rms (per~cent) & \\ \hline

1.66$\pm$0.04  &  715$\pm$12  &  98$\pm$50  &  2.0$\pm$0.4  &  985$\pm$7  &
 62$\pm$31 & 1.6$\pm$0.3 & 0.93 / 51 \\

1.79$\pm$0.04 &  715$^a$ & 98$^a$  & $<$1.4 &  985$^a$ & 62$^a$ &  $<$1.1 &
0.99 / 57 \\

  1.91$\pm$0.04 & 715$^a$ & 98$^a$ & $<$1.2 & 985$^a$ & 62$^a$ & $<$0.4 
 & 1.22 / 57 \\

\hline
\end{tabular}
\\
\noindent $^a$Fixed parameter.
\end{flushleft}
\end{table*}

\begin{table*}
\begin{flushleft}
\caption{Best-fitting values from energy resolved power spectra from the NB
  and FB.}
\label{tab:enefits}
\begin{tabular}{ccccccc} \hline 
Energy Range  & \multicolumn{2}{c}{VLFN} &  \multicolumn{2}{c}{Peaked Noise
rms~(per~cent)} & $\chi^{2}_{\nu}$/dof \\  (keV) & Index & rms~(per~cent) &
LFN/FBN & High Freq. Peak \\ \hline

Normal Branch \\

2.0--40  & 1.44$\pm$0.04 &  1.35$\pm$0.03 &  2.48$\pm$0.17 &  3.2$\pm$0.4 &
1.30 / 76 \\

2.0--5.1 & 1.10$\pm$0.12 & 0.85$\pm$0.03  & 2.11$\pm$0.07 & $<$3.0 & 1.50 /
81 \\

5.1--6.5 & 1.55$\pm$0.15 & 1.41$\pm$0.08  & 3.20$\pm$0.08 & $<$3.6 & 0.82 /
81 \\

6.5--8.7 & 1.68$\pm$0.12 & 1.74$\pm$0.10  & 3.61$\pm$0.07 & $<$4.1 & 1.44 /
81 \\

8.7--40  & 1.52$\pm$0.06 &  2.24$\pm$0.06 &  2.74$\pm$0.13 &  4.4$\pm$0.3 &
1.08 / 80 \\

Flaring Branch \\

2.0--40    &   1.72$\pm$0.02    &   3.25$\pm$0.05    &    2.98$\pm$0.04   &
1.0$^{+0.5}_{-0.2}$ & 1.28 / 76 \\

2.0--8.7 & 1.68$\pm$0.02 & 2.93$\pm$0.04  & 2.83$\pm$0.01 & $<$1.0 & 1.03 /
81 \\

5.1--40  & 1.73$\pm$0.02 &  3.89$\pm$0.06 &  3.45$\pm$0.06 &  1.3$\pm$0.1 &
1.31 / 80 \\

2.0--5.1 & 1.66$\pm$0.02 & 2.42$\pm$0.04  & 2.40$\pm$0.03 & $<$1.1 & 1.00 /
81 \\

5.1--6.5 & 1.69$\pm$0.03 & 3.22$\pm$0.06  & 3.14$\pm$0.04 & $<$1.1 & 1.26 /
81 \\

6.5--8.7 & 1.70$\pm$0.02 & 3.74$\pm$0.07  & 3.49$\pm$0.03 & $<$2.0 & 0.95 /
81 \\

8.7--40 & 1.77$\pm$0.02  & 4.71$\pm$0.09 & 3.78$\pm$0.03 &  $<$2.0 & 1.20 /
81 \\

\hline
\end{tabular}
\end{flushleft}
\end{table*}

\subsection{Low frequency noise/flaring branch noise peak
\label{sec:lfnfbnp}}

\begin{figure}
\begin{center}
\epsfig{file=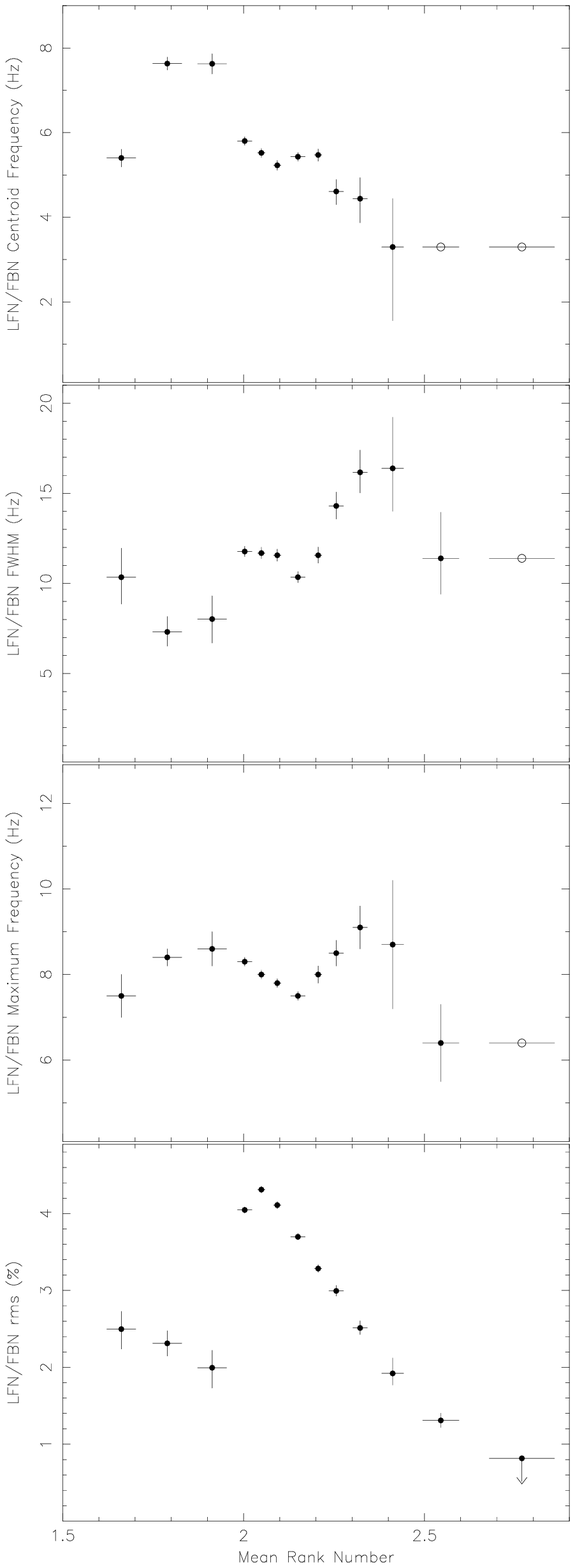,height=225mm,angle=0}
\caption{Low frequency noise/flaring branch noise centroid frequency, FWHM,
  $\nu_{\mathrm{max}}$, and fractional rms  amplitude as a function of rank
  number (2.0--40~keV).  The open circles indicate fixed values.}
\label{fig:fbn_val}
\end{center}
\end{figure}

In  Fig.~\ref{fig:fbn_val}   we  present  the   centroid  frequency,  FWHM,
$\nu_{\mathrm{max}}$, and fractional rms amplitude of the LFN/FBN peak (see
Fig.~\ref{fig:f_55_s}) as a function of S$_{z}$.

There is an increase in centroid frequency with movement down the NB, while
the FWHM, $\nu_{\mathrm{max}}$, and rms are consistent with being constant.
The mean $\nu_{\mathrm{max}}$ in  the NB was 8.2$\pm$0.2~Hz.  With movement
into  the  FB  the  centroid  frequency decreases  abruptly,  and  the  rms
increases  sharply.   There  is  no  significant change  in  the  FWHM  and
$\nu_{\mathrm{max}}$.  Overall as  the source moves up the  FB the centroid
frequency decreases while the  FWHM increases.  The $\nu_{\mathrm{max}}$ of
the peak  decreases as the object  moves from S$_{z}$=2.05 to  2.15, and it
then  increases   with  further   movement  up  the   FB.   We   note  that
$\nu_{\mathrm{max}}$ varies  less than the  centroid frequency. The  rms is
strongest at  S$_{z}$=2.05, and decreases until it  becomes undetectable in
the top half of the FB.

In the NB the  peak is strongest in 6.5--8.7~keV band, and  in the FB it is
strongest  in  the  highest  energy  band  (Table~\ref{tab:enefits}).   The
overall phenomenology  suggests that this peak  is LFN in the  NB, and then
evolves into, or is replaced by, FBN as the source moves into the FB.  This
interpretation is discussed further in Section~\ref{sect:lfnafbn}.

\begin{figure}
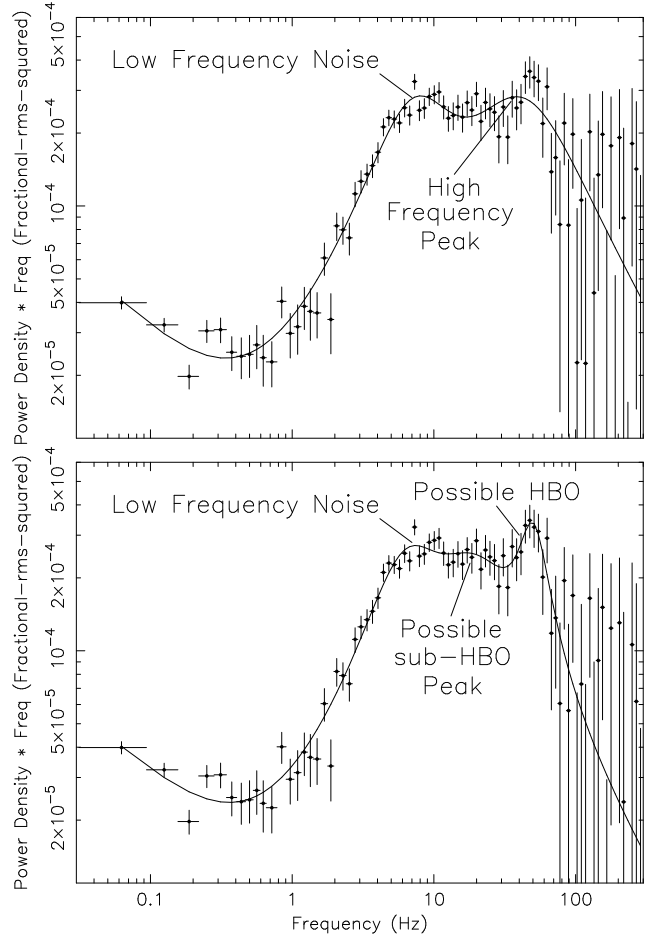

\begin{center}
\epsfig{file=MC390_fig4_top.ps,  height=84mm, angle=270}
\epsfig{file=MC390_fig4_bottom.ps, height=84mm, angle=270}
\caption{Power spectrum (2.0--40~keV) from the uppermost part of the NB
  (S$_{z}$=1.66), with Power Density$\times$Frequency on the vertical axis.
  The best-fitting  two- and three-Lorentzian  models are shown in  the top
  and bottom panels respectively.}
\label{fig:f_09_s_2lore}
\end{center}
\end{figure}

\subsection{High frequency peak \label{sec:hfp}}

\subsubsection{High frequency peak in the normal branch \label{sec:hfpitnb}}

In the  NB, the high frequency  peak (see Fig.~\ref{fig:f_55_s}  top) had a
centroid frequency in the range  11--22~Hz, and it exhibited no significant
change  in  its  centroid   frequency  or  fractional  rms  amplitude  (see
Table~\ref{tab:powfits2}).  There is marginal evidence of a decrease in the
FWHM with increasing rank number.   In energy resolved power spectra in the
NB   we   detected  the   peak   only   in   the  8.7--40~keV   band   (see
Table~\ref{tab:enefits}).

In  Fig.~\ref{fig:f_09_s_2lore} we  present the  2.0--40~keV  average power
spectrum from  the uppermost  part of the  NB, with S$_{z}$=1.66.   We used
power-density-times-frequency on  the vertical axis because  this shows the
power spectral  features more clearly.   A two-Lorentzian fit to  the power
spectrum  (Fig.~\ref{fig:f_09_s_2lore}~top;  Table~\ref{tab:powfits2}) gave
$\chi^{2}_{\nu}$/dof=1.37/76.    The    probability   of   exceeding   this
$\chi^{2}_{\nu}$  is   0.0178,  so  the  two-Lorentzian   fit  is  formally
satisfactory.   But $\chi^{2}_{\nu}$=1.37 is  still rather  high. Moreover,
there is a clear excess of power  in the range 45--60~Hz, which is where we
would  expect  to find  a  horizontal  branch  oscillation.  Therefore,  we
thought it  was worthwhile  to explore  the option of  fitting the  NB data
using three Lorentzians, even though  the addition of another Lorentzian to
the model is not statistically justified.

We used a three-Lorentzian model to fit the S$_{z}$=1.66 power spectrum and
obtained $\chi^{2}_{\nu}$/dof=1.16/73.   An F-test indicated  a 3.1$\sigma$
improvement to  the fit.  The best-fitting three-Lorentzian  model is shown
in Fig.~\ref{fig:f_09_s_2lore}~(bottom).  We also fitted the power spectrum
corresponding   to  S$_{z}$=1.79   in   the  same   fashion  and   obtained
$\chi^{2}_{\nu}$/dof=1.02/73.    An    F-test   indicated   a   3.5$\sigma$
improvement  to  the  fit.   The  best-fitting  values  from  both  of  the
three-Lorentzian fits are presented in Table~\ref{tab:powfits3}.

We  also fitted  the S$_{z}$=1.91  power spectrum  with  a three-Lorentzian
model and obtained  $\chi^{2}_{\nu}$/dof=1.06/73.  An F-test indicated this
to be a 1.4$\sigma$ improvement.  It  is not surprising that, in this power
spectrum, the three-Lorentzian model was not a significant improvement over
the two-Lorentzian model.  There is no obvious excess of power in the range
45--60~Hz,  in contrast  to  the  excess power  seen  at S$_{z}$=1.66.   In
Table~\ref{tab:powfits3} we show again, for comparison with the values from
the three-Lorentzian fits at S$_{z}$=1.66 and 1.79, the best-fitting values
from a two-Lorentzian fit to the S$_{z}$=1.91 power spectrum.

Through  comparison with  the other  Z sources,  in particular  Sco~X-1 and
GX~17+2 (e.g.,  Wijnands \& van~der~Klis \cy{wv99}  and references therein;
Homan et~al. \cy{hvj02}), we may  identify the three Lorentzians as: LFN, a
possible sub-HBO,  and a possible HBO.   A detailed comparison  is given in
Section~\ref{sec:tmfnbd}.

If we assume  we have detected three peaked  noise features at S$_{z}$=1.66
and 1.79,  and only  two features at  S$_{z}$=1.91, then, as  Sco~X-2 moves
down the NB,  the peak representing the possible  HBO disappears, while the
possible  sub-HBO  remains  strong  (see  Table~\ref{tab:powfits3}).   With
movement down the  NB, the $\nu_{\mathrm{max}}$ of the  possible sub-HBO is
consistent with remaining constant.

Even though an F-test indicated that, at S$_{z}$=1.66, the three-Lorentzian
model  was a  3.1$\sigma$  improvement over  the  two-Lorentzian model,  we
stress again that the two-Lorentzian model was formally a satisfactory fit.
Therefore, even though  we have investigated the use  of a three-Lorentzian
model,  we  cannot  conclude  we  have  detected an  HBO.   Since  we  have
\emph{not\/} detected an  HBO, we can measure an upper  limit by fixing the
centroid frequency and  FWHM at the values found  in, for example, Sco~X-1.
In that object, the centroid frequency was $\sim$45~Hz, and the FWHM was in
the   range   $\sim$8--40~Hz   (van~der~Klis  et~al.    \cy{vwh97}).    The
2.0--40~keV upper limits of an HBO  at S$_{z}$=1.66, with the FWHM fixed at
10, 20, 30 and 40~Hz, were 1.0, 1.6, 1.9 and 2.1~per~cent, respectively.

\begin{table*}
\begin{flushleft}
\caption{Best-fitting values from the three-Lorentzian fits to the
  S$_{z}$=1.66 and  1.79 power spectra,  and the two-Lorentzian fit  to the
S$_{z}$=1.91 power spectrum (2.0--40~keV).}
\label{tab:powfits3}
\setlength{\tabcolsep}{3pt}
\begin{tabular}{cccccccccccccc} \hline 
 S$_{z}$      &     \multicolumn{4}{c}{Low      Frequency      Noise}     &
 \multicolumn{4}{c}{Possible sub-HBO}  & \multicolumn{4}{c}{Possible HBO} &
 $\chi^{2}_{\nu}$/dof \\ & Freq &  FWHM & $\nu_{\mathrm{max}}$ & rms & Freq
 & FWHM & $\nu_{\mathrm{max}}$ & rms & Freq & FWHM & $\nu_{\mathrm{max}}$ &
 rms \\ & (Hz) & (Hz) & (Hz) & (per~cent) & (Hz) & (Hz) & (Hz) & (per~cent)
 & (Hz) & (Hz) & (Hz) & (per~cent) \\ \hline

1.66$\pm$0.04  &  5.1$\pm$0.3  &  8$\pm$2  & 6.5$\pm$0.7  &  2.2$\pm$0.4  &
13$\pm$5  & 29$\pm$10  & 19$\pm$5  & 2.3$\pm$0.7  & 48$\pm$2  &  26$\pm$8 &
50$\pm$2 & 1.5$\pm$0.3 & 1.16 / 73 \\

1.79$\pm$0.04  & 7.3$\pm$0.2 &  7.1$\pm$0.8 &  8.1$\pm$0.3 &  2.4$\pm$0.2 &
 16$\pm$2 &  16$\pm$6 &  18$\pm$2 & 1.9$\pm$0.5  & 46$\pm$10 &  59$\pm$17 &
 55$\pm$10 & 2.1$\pm$0.4 & 1.02 / 73 \\

1.91$\pm$0.04  & 7.6$\pm$0.2 &  8.0$\pm$1.3 &  8.6$\pm$0.4 &  2.1$\pm$0.2 &
14$\pm$4  & 32$\pm$6  &  21$\pm$3 &  2.4$\pm$0.3  & 46$^{a}$  & 59$^{a}$  &
55$^{a}$ & $<$1.5 & 1.09 / 76 \\

\hline
\end{tabular}
\\
\noindent $^a$Fixed parameter.
\end{flushleft}
\end{table*}

\subsubsection{High frequency peak in the flaring branch \label{sec:hfpitfb}}

We   initially   fitted  the   average   FB   spectrum  (2.0--40~keV;   see
Fig.~\ref{fig:f_55_s}~bottom)  with  a  power-law  and  single  Lorentzian,
representing   very  low   frequency  noise   and  flaring   branch  noise,
respectively, which gave  $\chi^{2}_{\nu}$/dof=1.67/79.  The probability of
exceeding  that $\chi^{2}_{\nu}$ is  0.00017, indicating  an unsatisfactory
fit at the 99.98~per~cent significance level.  Therefore, we added a second
Lorentzian  to  the model  and  obtained $\chi^{2}_{\nu}$/dof=1.28/76.   An
F-test  indicated this to  be a  4.1$\sigma$ improvement  to the  fit.  The
best-fitting centroid  frequency, FWHM and fractional rms  amplitude of the
additional   peak   were   54$^{+11}_{-45}$~Hz,  96$^{+69}_{-34}$~Hz,   and
1.0$^{+0.5}_{-0.2}$~per~cent,      respectively.       The     best-fitting
two-Lorentzian    model     is    shown     as    a    solid     line    in
Fig.~\ref{fig:f_55_s}~bottom.

We also detected the high  frequency peak in the 5.1--40~keV power spectrum
from    the    FB,   with    an    rms    of   1.3$\pm$0.1~per~cent    (see
Table~\ref{tab:enefits}).   The   nondetection  of  this   feature  in  the
2.0--8.7~keV  band, with an  upper limit  of 1~per~cent,  shows that  it is
genuinely hard.

\subsection{Very low frequency noise}

\begin{figure}
\begin{center}
\epsfig{file=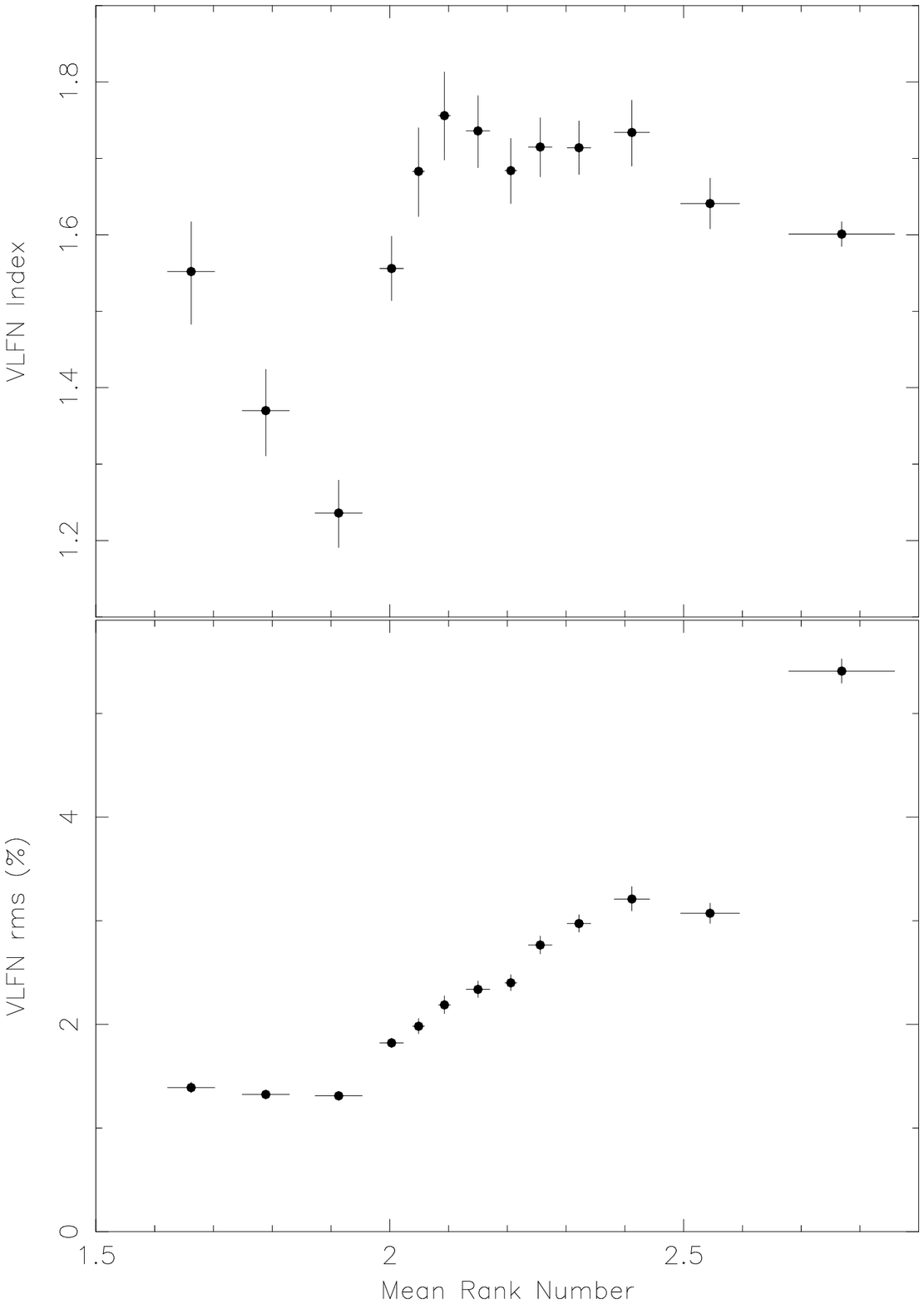,width=84mm,clip=}
\caption{VLFN index and fractional rms amplitude as a
function of rank number (2.0--40~keV).}
\label{fig:vlfn_ind}
\end{center}
\end{figure}

Very low  frequency noise was detected in  all regions of the  Z track.  In
Fig.~\ref{fig:vlfn_ind} we  present the power-law index  and fractional rms
amplitude of the VLFN as a  function of S$_{z}$.  The NB/FB vertex seems to
be a critical  position.  The index becomes less steep  as the source moves
down the  NB, and then abruptly  becomes steeper again at  the vertex.  The
index   remains  roughly   constant   with  movement   up   the  FB   until
S$_{z}$$\sim$2.5, after which  it decreases slightly.  In the  FB, the mean
value of the index is  1.70$\pm$0.02.  The rms is consistent with remaining
constant in  the NB, and then  increases abruptly at the  vertex.  The VLFN
then continues to  strengthen with movement up the FB,  and is strongest in
the highest energy band.

\subsubsection{VLFN and motion through the Z track \label{sect:vamttzt}}

We constructed an average power spectrum from the rank interval 1.87--1.88,
which  corresponds to the  vertical part  of the  NB in  the HID.   In this
region of  the NB there  is no associated  change in intensity  with motion
along the Z  track.  Therefore, if VLFN is  associated purely with movement
through the  track, then it  should be absent  in this power  spectrum.  We
found   VLFN  with   a  power-law   index  and   rms  of   1.1$\pm$0.2  and
1.1$\pm$0.1~per~cent,   respectively.   These   are  consistent   with  the
best-fitting  values of  the  VLFN found  in  other parts  of  the NB  (see
Table~\ref{tab:powfits2}).

\begin{figure}
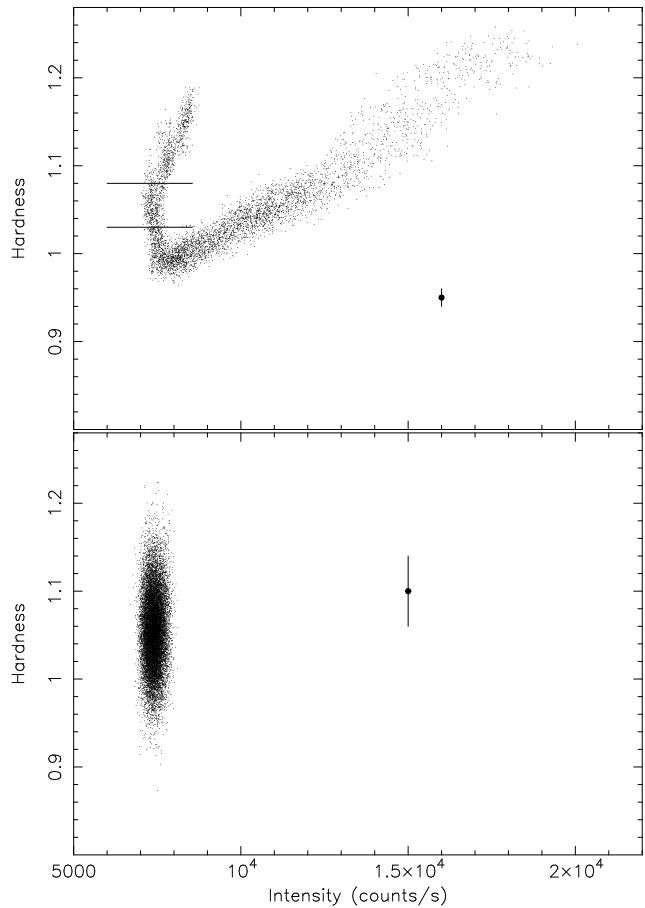

\begin{center}
\epsfig{file=MC390_fig6_top.ps,   height=84mm,angle=270}
\epsfig{file=MC390_fig6_bottom.ps, height=84mm,angle=270}
\caption{Hardness-intensity diagrams of single bit and event mode data. 
The  hardness  is the  counting  rate  ratio  between the  8.7--40~keV  and
6.5--8.7~keV bands,  and the  intensity is the  counting rate in  the range
2.0--40~keV.   The top  panel shows  all  data, using  32~s averages.   The
bottom panel shows  NB data in the hardness  range 1.03--1.08, indicated by
the horizontal lines in the  top panel, using 1~s averages.  Representative
error bars are shown.}
\label{fig:hid_32s}
\end{center}
\end{figure}

The detection of VLFN in the vertical  part of the NB suggests that VLFN is
not due solely  to movement along the track.  However,  it is possible that
within each stretch of data  over which Fourier transforms were calculated,
the source  moves a substantial distance  away from the  vertical region of
the NB,  by moving up  and down the  NB, or even  into the FB,  and thereby
producing       associated       changes       in      intensity.        In
Fig.~\ref{fig:hid_32s}~(top),  we  present the  HID  constructed from  32~s
averages     of    single     bit    and     event    mode     data.     In
Fig.~\ref{fig:hid_32s}~(bottom) we have plotted, in  a HID, all of the data
points in  the vertical part of  the NB (hardness  interval 1.03--1.08; see
Fig.~\ref{fig:hid_32s}~top) using  a time resolution of 1~s.   If there was
substantial movement up and down the  NB within each 32~s interval, then we
would expect the 1~s points to exhibit an arc-like distribution, reflecting
the shape of the NB.   We observed a vertical distribution, suggesting that
the changes in intensity comprising the VLFN are not due purely to movement
along the Z track.  However, we caution that the error bars are large.

A more sensitive way of  investigating the correlation between hardness and
intensity is  to use  cross spectra (see  Section~\ref{sect:iovlfn}).  With
phase measurements,  we can  determine if the  variability at  a particular
frequency is not entirely due to movement along the Z track.  In the FB and
the upper part of the NB,  where the track exhibits an increase in hardness
with increasing intensity  (see Fig.~\ref{fig:hid_32s}), movement along the
track must produce a phase of 0$\degr$.  In the lower part of the NB, where
the  branch exhibits  a  decrease in  hardness  with increasing  intensity,
movement along the track must produce a phase of 180$\degr$.

In Table~\ref{tab:phases}, we  present the phases from the  256~s, 32~s and
4~s intervals.  We only show phases  for the frequencies at which there was
a  detection  of  a  correlation  between hardness  and  intensity  at  the
90~per~cent  confidence  level.  The  upper  NB  corresponds  to data  with
hardness $>$1.06, and the lower NB to data in the hardness range 1.0--1.06.
For the FB,  we used data with intensity  $>$9000~counts~s$^{-1}$.  We also
present  the significance  of  the  detection of  the  correlation at  each
frequency, as measured from both the analytical and Monte Carlo method, and
rounded to the nearest one-hundredth of a per~cent.  Note that $>$3$\sigma$
detections were generally only obtained in the FB.

Since we wish to investigate VLFN,  we need to determine the frequencies at
which VLFN  is the primary influence  on the phases.   Above $\sim$2~Hz the
variations in  intensity are dominated  by LFN/FBN, which is  stronger than
the VLFN at those frequencies;  at lower frequencies the effects of LFN/FBN
may still be  present.  At frequencies where LFN/FBN  dominates, the phases
are characterised by non-zero values (see Table~\ref{tab:phases}); however,
we cannot, of course, assume that  every observation of a non-zero phase is
due to LFN/FBN.  If we observe,  at a particular frequency, a phase that is
consistent with being  zero, then we can conclude  that the phases observed
at lower frequencies must be due  to VLFN.  Any non-zero phases observed at
higher frequencies may  be due to either VLFN or LFN/FBN.   In the upper NB
and   FB,   all   phases    in   the   range   $\simeq$0.004--0.34~Hz   and
$\simeq$0.004--1.2~Hz,  respectively, are consistent  with being  zero.  In
the lower  NB: there  is a non-zero  phase at $\simeq$0.004~Hz;  the phases
from $\simeq$0.016 to $\simeq$0.09~Hz  are consistent with being zero; and,
there are non-zero phases at  frequencies in the range 0.1875--3.75~Hz.  We
shall, therefore,  use 0.1875~Hz as our  very rough estimate  of the lowest
frequency  at which  the phases  are primarily  influenced by  LFN/FBN, and
thereby conservatively restrict our study  of VLFN to frequencies less than
0.1875~Hz.

In the  FB and  upper NB,  the phases are  consistent with  being 0$\degr$,
indicating that the variations in intensity may be due entirely to movement
along the  Z track.  We combined the  256~s interval FB data  into a single
frequency bin,  with a mean  frequency of $\simeq$0.033~Hz, and  measured a
phase of $-$3\fdg 1$\pm$1\fdg 4, which is also consistent with zero.

In the  lower NB,  the phase at  $\simeq$0.004~Hz is consistent  with being
180$\degr$; therefore,  at this frequency,  the intensity and  hardness are
\emph{negatively\/}     correlated.      In     the     frequency     range
$\simeq$0.016--0.094~Hz,  the phases  are consistent  with  being 0$\degr$;
therefore,  in  this  frequency  range,  the  intensity  and  hardness  are
\emph{positively\/} correlated.  Since the Z track in the lower NB exhibits
a negative correlation between  hardness and intensity, the movement giving
rise to  the variations in  intensity in the  range $\simeq$0.016--0.094~Hz
must  have a  cross-track component.   And,  since we  observed a  negative
correlation  between  intensity   and  hardness  at  $\simeq$0.004~Hz,  the
cross-track component must become more prominent at higher frequencies.

It is  important to note again  the limitations of  our measurements.  Only
about  half of  our  detections are  at  the $>$3$\sigma$  level.  This  is
particularly a problem  in the lower NB, where we  have only $\sim$15~ks of
data.

\begin{table*}
\begin{flushleft}
\caption{Phases from average cross spectra from
the upper NB, lower NB, and FB.}
\label{tab:phases}
\begin{tabular}{ccccccc} \hline 
Frequency  &  \multicolumn{6}{c}{Phase  ($\degr$)  (per~cent  significance:
analytical/Monte   Carlo)}  \\  (Hz)   &  \multicolumn{2}{c}{Upper   NB}  &
\multicolumn{2}{c}{Lower NB} & \multicolumn{2}{c}{FB} \\ \hline

0.00390625 & 6$\pm$7 & (100/100)  & 174$\pm$19 & (94.56/94.80) & $-$4$\pm$2
  & (100/100) \\

0.0078125 & 8$\pm$15 & (99.95/99.99) & - & - & $-$3$\pm$3 & (100/100) \\

0.01171875 & $-$9$\pm$16  & (99.67/99.68) & - & -  & $-$2$\pm$3 & (100/100)
  \\

0.015625 &  $-$26$\pm$14 & (99.76/99.84)  & $-$57$\pm$26 &  (91.40/91.96) &
$-$4$\pm$3 & (100/100) \\

0.01953125 & $-$33$\pm$18 & (98.83/98.78) &  - & - & $-$1$\pm$4 & (100/100)
  \\

0.0234375 & 27$\pm$16 & (99.10/99.27) & - & - & 0$\pm$5 & (100/100) \\

0.02734375 & 6$\pm$29 & (91.28/91.74) & - & - & $-$3$\pm$6 & (100/100) \\

0.03125$^a$ & 38$\pm$25 & (91.67/91.43) & - & - & 9$\pm$7 & (100/100) \\

0.03125$^b$  &  20$\pm$8  &  (100/100)  & $-$48$\pm$21  &  (97.43/97.53)  &
$-$1$\pm$2 & (100/100) \\

0.03515625 & - & - & - & - & $-$2$\pm$8 & (100/100) \\

0.0390625 & 5$\pm$22 & (97.36/97.38) & - & - & $-$12$\pm$8 & (100/100) \\

0.04296875  & 28$\pm$21 &  (96.37/96.30) &  $-$15$\pm$25 &  (90.62/90.78) &
8$\pm$9 & (100/100) \\

0.046875 & - & - & - & - & $-$13$\pm$8 & (100/100) \\

0.05078125 & 14$\pm$22 & (91.91/92.10) & - & - & $-$4$\pm$8 & (100/100) \\

0.0546875 & 21$\pm$21 & (94.90/94.77) & - & - & 3$\pm$9 & (100/100) \\

0.05859375 & - & - & - & - & 3$\pm$8 & (100/100) \\

0.0625$^a$ & - & - & - & - & 0 & (99.94/99.98) \\

0.0625$^b$ & 12$\pm$13 & (99.99/100) & - & - & $-$7$\pm$4 & (100/100) \\

0.09375  &  0$\pm$19  &  (99.11/99.18)  &  $-$8$\pm$20  &  (97.70/97.78)  &
$-$8$\pm$6 & (100/100) \\

0.125 & $-$16$\pm$20 & (97.38/97.37) & - & - & $-$12$\pm$6 & (100/100) \\

0.15625 & $-$53$\pm$20  & (98.43/98.41) & - & -  & $-$13$\pm$10 & (100/100)
  \\

0.1875 & - & - & 176$\pm$24 & (94.12/93.74) & $-$13$\pm$12 & (100/100) \\

0.21875 & $-$13$\pm$18 & (99.27/99.33) &  - & - & 15$\pm$18 & (99.55/99.59)
  \\

0.25$^b$ & - & - & - & - & - & - \\

0.25$^c$ & $-$25$\pm$22  & (96.34/96.26) & - & -  & $-$5$\pm$12 & (100/100)
  \\

0.28125 & - & - & - & - & $-$24$\pm$26 & (91.46/90.99) \\

0.3125 & - & - & - & - & - & - \\

0.34375 & $-$15$\pm$23 & (96.38/96.06) & 112$\pm$22 & (94.62/94.50) & - & -
  \\

0.375 & - & - & $-$177$\pm$28 & (90.55/90.73) & 6$\pm$21 & (97.43/97.82) \\

0.4062 & - & - & - & - & 28$\pm$20 & (98.03/97.82) \\

0.4375 & - & - & - & - & - & - \\

0.46875 & - & - & - & - & - & - \\

0.5$^b$ & - & - & - & - & - & - \\

0.5$^c$ & - & - & - & - & - & - \\

0.75 & 102$\pm$16 & (99.78/99.76) & - & - & - & - \\

1.0 & - & - & - & - & - & - \\

1.25 & - & - & - & - & 35$\pm$26 & (90.59/90.29) \\

1.5 & - & - & - & - & - & - \\

1.75 & $-$162$\pm$20 & (98.53/98.44) & - & - & -\\

2.0 & $-$178$\pm$25 & (91.79/91.25) & - & - & - & - \\

2.25 & - & - & - & - & - & - \\

2.5 & - & - & 167$\pm$24 & (94.48/94.44) & - & - \\

2.75 & - & - & 74$\pm$23 & (95.63/95.29) & - & - \\

3.0 & 80$\pm$25 & (92.61/92.55) & - & - & 110$\pm$25 & (92.46/92.29) \\

3.25 & - & - & - & - & - & - \\

3.5 & - & - & - & - & - & - \\

3.75 & - & - & $-$131$\pm$18 & (99.19/99.33) & - & - \\

4.0 & - & - & - & - & - & - \\

\hline
\end{tabular}
\\
\noindent $^a$256~s intervals.\\$^b$32~s intervals.\\$^c$4~s intervals. 
\end{flushleft}
\end{table*}

\section{Discussion \label{sect:disc}}

Using Zhang~et~al.'s~(\cy{zss98})  {\it RXTE \/} data, we  have carried out
the most comprehensive investigation  of the X-ray fast-time variability of
Sco~X-2 thus  far, as a function of  position on the Z  track.  Our primary
discoveries are:  the properties of the low  frequency noise/flaring branch
noise peak, and  the very low frequency noise,  exhibit abrupt changes with
movement from the  NB into the FB;  there is a peaked noise  feature with a
centroid frequency  in the range 11--54~Hz,  which, in the  NB, may contain
either one or two components; and, very low frequency noise is not entirely
due to motion along the Z track.

\subsection{Low frequency noise and flaring branch noise \label{sect:lfnafbn}}

The peak  we observed in the  NB and FB,  with a centroid frequency  in the
range 3.3--7.6~Hz, is not an N/FBO: NBOs are much narrower than the peak we
observed; the centroid frequency of  an FBO increases rapidly with movement
up the FB, in contrast to our peak; and FBOs are only present in the lowest
10--20~per~cent of  the FB, while the  feature in Sco~X-2  is present until
about  halfway up the  branch (see  also Section~\ref{sec:sx}).   Given the
presence of  kHz QPOs in  the upper NB,  and the fact  that the peak  has a
$\nu_{\mathrm{max}}$ in the range 7.5--9.1~Hz, which is consistent with the
LFN  found in  other Z  sources (e.g.,  GX~17+2, Homan  et~al.  \cy{hvj02};
Sco~X-1, Dieters \& van~der~Klis \cy{dv00}), we argue that we have detected
LFN in the NB.  In the FB, the  peak is similar to the FBN found earlier in
Sco~X-2 (e.g., Agrawal \&  Bhattacharyya \cy{ab01} and references therein),
and we identify it as such.

LFN  is a  hard phenomenon  (e.g., van~der~Klis  \cy{v95a}).   For example,
Jonker et~al. (\cy{jvw00})  found that the fractional rms  amplitude of the
LFN in GX~340+0  was more than a factor of three  greater in the 13--60~keV
band than it was in the 2--5~keV band.  In contrast, the LFN we observed is
strongest in the 6.5--8.7~keV band.  This difference suggests that maybe we
have \emph{not\/} observed LFN,  since different production mechanisms give
rise  to  different  energy  dependences  (e.g., compare  HBOs  with  NBOs,
van~der~Klis  \cy{v95a} and references  therein).  However,  the fractional
rms amplitude  spectrum of  NBOs, for example,  is known to  differ between
objects (e.g.,  van~der~Klis \cy{v95a}  and references therein;  Dieters \&
van~der~Klis  \cy{dv00}; Dieters  et~al.   \cy{dvk00}) and  LFN may  behave
similarly different per source.  This suggestion can be confirmed through a
sensitive  investigation of  the  S$_{z}$  dependence of  the  LFN that  we
detected, and by making a comparison with other Z sources.

In previous observations  of Sco~X-2 the properties of  the peaked noise in
the NB was found  to be similar to that seen in  the FB, suggesting that it
was the same feature in both branches (Agrawal \& Bhattacharyya \cy{ab01}).
We argue, for two reasons, that  FBN is different to LFN: the properties of
the LFN/FBN peak, as seen in Fig.~\ref{fig:fbn_val}, exhibit abrupt changes
at the NB/FB vertex; and the  energy dependence of the peak differs between
NB and FB. We note that  according to the `unified model' (Psaltis, Lamb \&
Miller \cy{plm95}),  the inferred mass-accretion  rates are expected  to be
sub-Eddington in the NB, and super-Eddington in the FB, suggesting that the
mechanism producing the  variability in the NB may be  different to that in
the FB.   However, the  properties of the  N/FBO, for example,  also differ
between  NB and  FB,  with the  changes  occurring near  the vertex  (e.g.,
Dieters \& van~der~Klis  \cy{dv00}), and the NBO and  FBO are still thought
to be  related.  Likewise, there  remains the possibility that  LFN evolves
into FBN.

Our observations of  FBN support the conclusions reached  by O'Neill et~al.
(\cy{oks01}), which were  detailed in Section~\ref{sec:sx}.  In particular,
we note again the similarity between FBN  and the broad peak seen in the FB
of Cyg~X-2  at low overall intensities (Kuulkers,  Wijnands \& van~der~Klis
\cy{kwv99}; see  also O'Neill  et~al. \cy{oks01}; Agrawal  \& Bhattacharyya
\cy{ab01}).  The  feature in Cyg~X-2  was fitted with a  cut-off power-law,
with a  $\nu_{\mathrm{max}}$ of 10$\pm$3~Hz.   This is consistent  with the
FBN we observed.   The peak in Cyg~X-2  was not observed in the  NB, with a
fractional rms amplitude upper limit  of 1.0~per~cent, compared with the FB
rms of 2.7$\pm$0.2~per~cent.  An investigation of the S$_{z}$ dependence of
the Cyg~X-2  peak will reveal whether or  not it is the  same phenomenon as
FBN.  Furthermore, a stringent upper limit on the rms of the peak in the NB
would support our suggestion that FBN is not the same as LFN.

We  now compare  the  LFN and  FBN we  observed  in Sco~X-2,  with the  LFN
observed  in the  NB  and FB  of  GX~17+2 (Homan  et~al. \cy{hvj02}).   The
$\nu_{\mathrm{max}}$ of  the LFN in  the NB of  Sco~X-2 is generally  a few
hertz lower than that seen in the  NB of GX~17+2, and it is consistent with
the $\nu_{\mathrm{max}}$ observed  in the HB.  There is  no abrupt decrease
in the centroid  frequency and increase in the  fractional rms amplitude of
the LFN in GX~17+2, with movement  from the NB into the FB, suggesting that
it   may  not   be  the   same  as   Sco  X-2   LFN/FBN.    However,  Homan
et~al. (\cy{hvj02})  stated that  their best-fitting values  of LFN  in the
lower  NB and  lower FB  were  probably influenced  by the  N/FBO that  was
present, so the  true LFN phenomenology may have  been masked.  The overall
behaviour of $\nu_{\mathrm{max}}$ in the FB is similar in both objects, and
the decrease  in the fractional rms  amplitude of the LFN  in GX~17+2, with
movement up the  FB, is remarkably similar to FBN.  In  contrast to the FBN
we observed, the  LFN in GX~17+2 was detected as  high as about 85~per~cent
of the  way up the  FB.  But, we  note again that Agrawal  \& Bhattacharyya
(\cy{ab01}) reported the detection of peaked  noise in the FB of Sco~X-2 at
about 80~per~cent of the way up the FB.

In addition to  GX~17+2, a close examination of the LFN  in Sco~X-1 is also
desirable.   Sco~X-1  has  been  observed  by {\it  RXTE},  with  2--60~keV
counting  rates  of  $\sim$10$^{5}$~counts~s$^{-1}$,  and  these  data  are
currently  available  in the  archive  (e.g.,  Yu,  van~der~Klis \&  Jonker
\cy{yvj01}).   Of course,  more data  on Sco~X-2  are also  required before
sufficiently detailed comparisons can be made.

\subsection{High frequency peak in the normal branch}

In  the NB,  we detected  a  high frequency  peaked noise  feature, with  a
centroid frequency in the range  11--22~Hz.  This feature contains power at
frequencies  where  we  would  expect  to  find  high  frequency  noise,  a
horizontal branch  oscillation, and/or a  sub-horizontal branch oscillation
(see Section~\ref{sec:zaasfv}).  We showed that the NB power spectrum could
be fitted satisfactorily with  a two-Lorentzian model, using one Lorentzian
for  the low  frequency noise,  and one  for the  high frequency  peak (see
Section~\ref{sec:hfp}).  However,  at S$_{z}$=1.66,  there is an  excess of
power in  the frequency  range where we  would expect  to find an  HBO (see
Fig.~\ref{fig:f_09_s_2lore}).  The addition of a third Lorentzian component
significantly improved the fit.  Therefore, we shall consider both two- and
three-Lorentzian models for fitting the NB data.

\subsubsection{Two-Lorentzian model for normal branch data}

The high frequency peak, if  considered as a single Lorentzian, resembles Z
source HFN.  The $\nu_{\mathrm{max}}$ we found is consistent with that seen
in other Z  sources (e.g., Hasinger \& van~der~Klis  \cy{hv89}), and it was
stronger  at higher energies,  which is  typical of  HFN (e.g.,  Dieters \&
van~der~Klis \cy{dv00}).  However, there is  some doubt as to the origin of
HFN:  it may  be an  instrumental  effect from  \emph{EXOSAT\/} (Berger  \&
van~der~Klis \cy{bv94}).  Homan et~al.   (\cy{hvj02}) did not detect HFN in
a  large  \emph{RXTE\/}  dataset   on  GX~17+2,  in  contrast  to  previous
observations made by \emph{EXOSAT\/}.  They suggested that in the HB in the
\emph{EXOSAT\/}  observations,  where the  HFN  was  too  strong to  be  an
instrumental effect,  the feature  being fitted in  power spectra  may have
actually been the sub-HBO peak.

Is then  the high frequency peak,  taken as a single  Lorentzian, a sub-HBO
peak?   At  the  top  of  the  observed  NB, the  LFN  we  detected  had  a
$\nu_{\mathrm{max}}$ of  7.5$\pm$0.5~Hz.  The centroid  frequency, FWHM and
$\nu_{\mathrm{max}}$  of   the  high  frequency   peak  were  22$\pm$11~Hz,
70$\pm$11~Hz and 41$\pm$7~Hz, respectively.  In comparison, when the LFN in
GX~17+2 had a $\nu_{\mathrm{max}}$ of  $\sim$7.5~Hz, its sub-HBO peak had a
centroid  frequency,  FWHM,  and $\nu_{\mathrm{max}}$  of  25.2$\pm$1.3~Hz,
21$\pm$7~Hz and  27$\pm$2~Hz, respectively (Homan  et~al. \cy{hvj02}).  The
high frequency peak  in Sco~X-2 is at the expected  centroid frequency of a
sub-HBO, but there are some  marked differences: we would expect to clearly
detect an  HBO when a sub-HBO peak  is present; the high  frequency peak is
much  broader  than a  typical  sub-HBO, which  results  in  a much  higher
$\nu_{\mathrm{max}}$;    and,   with    movement   down    the    NB,   the
$\nu_{\mathrm{max}}$    of   the   high    frequency   peak    in   Sco~X-2
\emph{decreased\/} to  21$\pm$3, while in GX  17+2 the $\nu_{\mathrm{max}}$
of  the  sub-HBO  peak  \emph{increased\/}  to a  maximum  of  $\sim$50~Hz.
Therefore, we  conclude that  the high frequency  peak, if considered  as a
single Lorentzian,  can not be identified  as a typical  manifestation of a
sub-HBO.

\subsubsection{Three-Lorentzian model for normal branch data
\label{sec:tmfnbd}} 

We now  consider the possibility that  the high frequency  peak consists of
two Lorentzian features.   At S$_{z}$=1.66, the peak could  be fitted using
two Lorentzians, representing a possible sub-HBO and HBO.  In the same part
of the NB, the lower kHz QPO is at 715$\pm$12~Hz.

We  find  that,   according  to  the  correlation  found   by  Wijnands  \&
van~der~Klis   (\cy{wv99}),   and   in   comparison   to   GX~17+2   (Homan
et~al. \cy{hvj02}), the frequencies of the LFN and possible sub-HBO peak we
observed in Sco~X-2 are consistent with expectations, assuming the presence
of an HBO at 48~Hz.

According to the correlation found by Psaltis~et~al. (\cy{pbv99}), when the
lower kHz QPO is at 715~Hz the HBO is inferred to have a centroid frequency
of about 60~Hz.  This is  somewhat higher than the 48~Hz centroid frequency
of our possible  HBO.  An HBO centroid frequency of  48~Hz corresponds to a
lower kHz QPO  frequency of about 600~Hz (Psaltis  et~al. \cy{pbv99}; Homan
et~al.   \cy{hvj02}).  The  HBO  is  Sco~X-1 has  only  been detected  with
centroid frequencies  in the range  $\sim$40--50~Hz, and Sco~X-1  also lies
off the Psaltis  et~al. (\cy{pbv99}) correlation; for example,  its HBO was
found  at $\sim$45~Hz, even  when the  lower kHz  QPOs was  at $\sim$875~Hz
(van~der~Klis et~al. \cy{vwh97}).  Therefore,  we speculate that Sco~X-2 is
similar to Sco~X-1 in this regard, and predict that future observations may
reveal an HBO in the same frequency range as seen in Sco~X-1.

There  remains   a  problem,  however,  with  the   interpretation  of  our
$\nu_{\mathrm{max}}$$\sim$20~Hz feature as being a Z source sub-HBO.  If we
assume  we have detected  an HBO  and sub-HBO  in the  NB power  spectra at
S$_{z}$=1.66 and 1.79, then, as Sco~X-2 moves down the NB, the sub-HBO peak
should  disappear before  the HBO  does.  This  effect is  clearly  seen in
GX~17+2, in which the sub-HBO  peak becomes undetectable in the middle part
of  the NB while  the HBO  is present  all the  way down  to the  FB (Homan
et~al. \cy{hvj02}).  In  contrast, we find that our  possible HBO component
disappears by  S$_{z}$=1.91, while  the possible sub-HBO  component remains
strong.   Furthermore, the $\nu_{\mathrm{max}}$  and centroid  frequency of
the possible sub-HBO at S$_{z}$=1.79 and 1.91, are significantly lower than
the maximum values observed in the NB in GX~17+2 (Homan et~al. \cy{hvj02}).
However, we note that the uncertainties on our measurements do allow for an
increase    in    the    possible    sub-HBO   centroid    frequency    and
$\nu_{\mathrm{max}}$.  Therefore, although  problematic, we cannot rule out
the possibility that we have observed a manifestation of a sub-HBO slightly
different to that seen in GX~17+2.

In summary,  if three  Lorentzians \emph{are\/} required  to fit  the power
spectrum from  the upper NB, we  conclude that: the frequency  of the lower
kHz QPO, relative to the centroid  frequency of the HBO, is similar to that
seen  in Sco~X-1 (van~der~Klis  et~al.  \cy{vwh97});  and the  feature with
$\nu_{\mathrm{max}}$$\sim$20~Hz may be a sub-HBO  peak, though it must be a
different manifestation to that seen in GX~17+2.

\subsection{Comparison with models}

Since  flaring   branch  noise  is  present   at  super-Eddington  inferred
mass-accretion rates,  where the accreting material  and outgoing radiation
must be separated (e.g., Lamb \cy{l91}), an adaptation of the photon bubble
model (Klein  et~al. \cy{kaj96a}) may  prove useful in explaining  it.  The
photon  bubble model  was originally  applied  to X-ray  pulsars, in  which
locally  super-Eddington accretion  takes place  onto the  polar caps  of a
strongly  magnetised  neutron  star  (Klein et~al.   \cy{kaj96a}).   Photon
bubbles are  formed in the settling  mound below the  accretion shock.  The
bubbles coalesce  to form larger bubbles,  and then rise up  and lose their
photons  through  diffusion  into  the  accretion  shock.   The  result  is
quasi-periodic variability  in the X-ray intensity, the  frequency of which
is primarily dependent upon the time  it takes for the bubbles to coalesce.
Modelling showed that QPOs could be  produced at frequencies from as low as
20~Hz to as high as 12000~Hz, along with a simultaneous power-law component
(Klein  et~al.   \cy{kaj96a};  Klein  et~al.  \cy{kja96b}).   It  would  be
interesting to see whether or not  a similar process, operating in either a
radial inflow  or in  the accretion funnel  onto the neutron  star magnetic
poles, could give rise to FBN.

In  the   magnetospheric  beat   frequency  model  for   horizontal  branch
oscillations  (e.g., Alpar  \& Shaham  \cy{as85}; Lamb  et~al.  \cy{lsa85};
Shibazaki \& Lamb  \cy{sl87}), the HBO is predicted to  be accompanied by a
noise component  at lower  frequencies, with both  features expected  to be
hard.  In the power spectra of  Z sources it is possible to identify either
the  LFN or  the sub-HBO  peak as  the noise  component (e.g.,  Wijnands \&
van~der~Klis \cy{wv99}).  The LFN that  we observed in the NB was strongest
at  intermediate   energies,  suggesting  that,  in  the   context  of  the
magnetospheric beat frequency model, either  LFN is not related to the HBO,
or  the feature  we  observed is  not  classical LFN.   We  have argued  in
Section~\ref{sect:lfnafbn} that we \emph{have\/} observed Z source LFN.  If
we are correct, then the sub-HBO peak, rather than LFN, might be identified
as the noise component that is expected to accompany the HBO.

In the relativistic precession model, the upper and lower kHz QPOs, and the
HBO,  are all  produced by  the same  blob orbiting  in the  accretion disc
(Stella \& Vietri \cy{sv98}; Stella  \& Vietri \cy{sv99}; Stella, Vietri \&
Morsink \cy{svm99}).  The model makes clear predictions of the correlations
between  those frequencies.   Unfortunately,  since we  did not  definitely
detect an  HBO, and the kHz  QPOs were only detected  in one part  of the Z
track, we cannot test those predictions.  And, our HBO upper limits are too
high to  draw any  meaningful conclusion about  its presence  or otherwise,
since weaker HBOs have been observed (e.g., in Sco~X-1; van~der~Klis et~al.
\cy{vwh97}).  If we assume that  we \emph{have\/} detected an HBO at 48~Hz,
with kHz QPOs at 715~Hz and 985~Hz, then the mass and spin frequency of the
neutron star can be inferred from the model.  Judging from fig.~1 in Stella
\& Vietri (\cy{sv99}), the inferred mass is $\sim$2~M$_{\sun}$; the other Z
sources have inferred masses in the range $\sim$1.8--2.2~M$_{\sun}$ (Stella
\& Vietri  \cy{sv99}).  The mass  of the neutron  star in Cyg~X-2  has been
dynamically  established as  1.78$\pm$0.23~M$_{\sun}$, which  is consistent
with  the inferred  masses  (Orosz \&  Kuulkers  \cy{ok99}).  Judging  from
fig.~1(b) in  Stella et~al. (\cy{svm99}), the inferred  spin frequency, for
neutron star equation  of state AU and an  assumed mass of 1.95~M$_{\sun}$,
is $\sim$600~Hz; the other Z sources have an inferred spin frequency in the
range $\sim$600--900~Hz (Stella et~al. \cy{svm99}).

In  the so-called transition  layer model  (e.g., Titarchuk,  Osherovich \&
Kuznetsov \cy{tok99}), Z source LFN  is the same phenomenon as atoll source
HFN, and the  sub-HBO peak is the same phenomenon  as atoll source 1--67~Hz
QPOs.  Both  of these phenomena are  due to oscillations  in the transition
layer  between the  disc and  the neutron  star.  The  HBO, the  HBO second
harmonic,  and  the  upper  kHz  QPO,  are  frequencies  at  which  a  blob
oscillates,  having  been thrown  out  of  the  transition layer  into  the
magnetosphere.  The  lower kHz  QPO is the  orbital frequency at  the inner
edge of the disc.  Recently, Wu (\cy{w01}) found good agreement between the
predictions of the  model and observations of Z  sources and atoll sources.
It is unclear how the fractional  rms amplitude of LFN is predicted to vary
with energy because the actual mechanism that produces that variability has
not been specified (Titarchuk \& Osherovich \cy{to99}).  Naturally any such
mechanism  that is  proposed  must  account for  the  energy dependence  we
observed  in Sco~X-2.  If  the transition  layer model  is correct,  we can
measure the angle $\delta$ between the  plane of the accretion disc and the
equator of  the magnetosphere (e.g.,  Titarchuk et~al. \cy{tok99}).   If we
assume an HBO frequency of 48~Hz, and use our observed kHz QPO frequencies,
then  we infer  $\delta$=5\fdg 6$\pm$0\fdg  3,  which is  within the  5\fdg
4--6\fdg 4 range found from the other Z sources (Wu \cy{w01}).

The sub-HBO  seen in Z  sources may be  interpreted as the  fundamental QPO
frequency, with  the HBO and its  second harmonic being  interpreted as the
second  and fourth  harmonics of  the sub-HBO  (Jonker  et~al. \cy{jvw00}).
Jonker et~al. (\cy{jvh02}) (also Jonker  \cy{j01}) found a broad feature in
GX~5$-$1 at  1.5 times the HBO  frequency, making it the  third harmonic of
the sub-HBO.   They suggested that a  warped disc with  a two-fold symmetry
can provide a mechanism which  produces odd harmonics that are broader than
the even  harmonics, and that this mechanism  could be used as  part of the
relativistic precession model.  The FWHM of the possible sub-HBO and HBO we
observed at  S$_{z}$=1.66 were 29$\pm$10~Hz  and 26$\pm$8~Hz, respectively.
The uncertainties here are too large to test the warped disc model.

\subsection{Very low frequency noise}

\subsubsection{The importance of studying VLFN \label{sec:tioavm}}

If  VLFN is  due  to changes  in  intensity that  are  associated with  the
movement of a source along its Z  track, then a model for VLFN will provide
at least part of the mechanism that gives rise to the Z track.

For example,  in the unified  model (e.g., Psaltis et~al.   \cy{plm95}) the
movement along the Z track is  driven by changes in the mass-accretion rate
which alter the  physical conditions in the region of  the neutron star and
inner  accretion  disc.   The  mass-accretion  rate changes  from  sub-  to
super-Eddington with movement from the  NB into the FB.  In this particular
context, a  model for VLFN will  provide the mechanism  that produces those
changes in the accretion rate.

Recently, van~der~Klis (\cy{v01}) suggested that position along the Z track
may not be related to  the actual mass-accretion rate.  Rather, he proposed
that rank number  is related to the ratio  $\eta$ between the instantaneous
accretion rate  through the disc, and  the long-term average  of that rate.
In this model, movement to a higher S$_{z}$ requires either: an increase in
the  accretion  rate;  or  for   the  accretion  rate  to  remain  constant
immediately following a decrease.   Conversely, movement to a lower S$_{z}$
occurs  either: when  the  accretion  rate decreases;  or  when it  remains
constant  immediately following  an increase.   Again,  if VLFN  is due  to
movement along the track, then the  details of this model must be such that
$\eta$ exhibits a power-law variability.

Furthermore, the abrupt  changes we observed in the  properties of the VLFN
with movement  through the  NB/FB vertex suggests  that any model  for VLFN
should  also   provide  information  about  the  physical   basis  of  that
transition.

\subsubsection{Observations of VLFN \label{sec:oov}}

We observed  an increase in the  fractional rms amplitude of  the VLFN with
increasing  S$_{z}$, and  it was  stronger  at higher  energies.  The  same
behaviour was seen  in the {\it Ginga \/}  observations of Sco~X-2 (O'Neill
et~al.   \cy{oks01}),   and  also  in  Sco~X-1   (Dieters  \&  van~der~Klis
\cy{dv00}).  However, our results are very different to those from the 1998
September and  October {\it  RXTE \/}  data on Sco~X-2,  in which  the VLFN
behaved somewhat erratically, and was,  on average, stronger in the NB than
in the FB (Agrawal  \& Bhattacharyya \cy{ab01}).  Homan et~al. (\cy{hvj02})
found that, in GX~17+2, the rms  of the VLFN did not increase smoothly with
movement up the FB; as the object moved away from the NB/FB vertex, the rms
decreased at  first, reaching a local  minimum at about  one-quarter of the
way  up  the FB,  and  then  increased  smoothly with  increasing  S$_{z}$.
However, the VLFN in GX~17+2 was still, on average, stronger in the FB than
in the  NB.  The reason for  the differences between our  results and those
from  the  1998  September  and   October  data  is  unclear.   Agrawal  \&
Bhattacharyya (\cy{ab01})  used 8~s intervals when  calculating their power
spectra,  which meant  their  lowest frequency  was  0.125~Hz, compared  to
0.0625~Hz in our  spectra; this may have produced  some differences between
the two  sets of results due  to interference, during  the fitting process,
from the peaked noise.

The power-law  index of the VLFN  we observed decreased  with movement down
the NB, abruptly  increased at the NB/FB vertex,  and then remained roughly
constant with movement up the FB,  decreasing slightly in the upper part of
the  FB; in the  FB, the  mean value  of the  index was  1.70$\pm$0.02.  In
contrast to  this, the index  of the VLFN  found in the 1998  September and
October {\it  RXTE \/} data exhibited  no dependence on S$_{z}$;  in the FB
the  index had  a mean  value  of 1.64$\pm$0.07  (Agrawal \&  Bhattacharyya
\cy{ab01}).   In  Sco~X-1,  the  power-law  index  \emph{increased\/}  with
movement down  the NB, from 1.2  to 1.7, although, similar  to our results,
was  constant  in  the  FB,  with  a  mean  of  1.76$\pm$0.04  (Dieters  \&
van~der~Klis \cy{dv00}).   In GX~17+2,  the index gradually  increased with
movement down the NB and reached  a maximum at about 10~per~cent of the way
up the  FB; it  then decreased  gradually with further  movement up  the FB
(Homan et~al.  \cy{hvj02}).  Of note,  however, is that the power-law index
in GX~17+2 reached a local  maximum of $\sim$1.6, at roughly three-quarters
of the way down the NB, and it then subsequently decreased to reach a local
minimum,  of  $\sim$0.8,  near  the  NB/FB vertex  (see  fig.~11  in  Homan
et~al. \cy{hvj02}). A similar effect has also been seen in GX~5$-$1 (Jonker
et~al. \cy{jvh02}),  and we  may have observed  the same thing  in Sco~X-2,
albeit with a much broader local minimum.

As discussed  in the previous  section, understanding VLFN is  important in
understanding the overall behaviour  of Z sources.  Therefore, the presence
of  vastly  different VLFN  phenomenologies  in  Sco~X-2, during  different
observations, is an important result  that needs to be clarified.  Wijnands
et~al.  (\cy{wvk97})  found that,  in Cyg~X-2, the  properties of  the VLFN
varied  as  the  overall  intensity  level  changed;  variations  in  other
components  of  the  fast-time  variability,  and  in  the  X-ray  spectral
properties, also  accompanied those changes.   The differences in  the VLFN
phenomenology in Sco~X-2 may be related to the same effect.

\subsubsection{VLFN and movement through the Z track
\label{sec:oovlamttzt}}

Dieters  \&  van~der~Klis  (\cy{dv00})  identified  VLFN  in  Sco~X-1  with
movement along the Z track,  while Homan et~al. (\cy{hvj02}) found that, in
GX~17+2, the speed of movement through  the Z track was not correlated with
VLFN  strength in  all parts  of the  track.  Wijnands  et~al. (\cy{wvk97})
similarly   investigated  the   VLFN  in   Cyg~X-2  at   different  overall
intensities.  They found  that the speed of movement  through the track was
highest  at  the  highest  overall  intensity level,  while  the  VLFN  was
strongest at  the intermediate level;  they, therefore, concluded  that the
VLFN in the NB could not solely be due to movement along the Z track.

We detected  VLFN in the  vertical part of  the NB.  Therefore, if  VLFN is
produced  solely by  intensity changes  that are  associated  with movement
along  the  track,  then,  within  each interval,  there  must  significant
movement along  the track to regions  above and below the  vertical part of
the NB,  such that there  are variations in  intensity.  The absence  of an
arc-like distribution of  the 1~s time resolution points  suggests there is
no  such  movement.   The  1~s  time-domain data  points  have  very  large
uncertainties  in   hardness,  so  we  used  cross   spectral  analysis  to
investigate the relationship between intensity and hardness.

In the  FB and upper  NB the variations  in the intensity and  hardness are
consistent with  being in phase with  each other.  This  is consistent with
the variations in  intensity being due only to movement  along the Z track,
though it does not rule out a cross-track component.

In  the lower  NB,  we found  that  at $\simeq$0.004~Hz  the variations  in
intensity were negatively correlated with the variations in hardness, which
is consistent with those variations also  being due to movement up and down
that part  of the NB.   However, at higher  frequencies, we found  that the
intensity  was positively  correlated  with hardness,  indicating that,  at
those  frequencies, the  variations in  intensity must  be  associated with
movement  that  is not  directly  along the  Z  track.   And, the  negative
correlation at $\simeq$0.004~Hz suggests that that cross-track component of
motion is more prominent at  higher frequencies.  We caution, however, that
these conclusions  must remain tentative  due to the low  significance, and
large uncertainties, of our phase measurements.

\subsubsection{Models for VLFN \label{sec:mfv}}

A model  has been proposed to  explain power-law variability  in black hole
candidates  (e.g.,  Mineshige, Takeuchi  \&  Nishimora \cy{mtn94}).   Their
model can  produce power-law fluctuations  in the mass-accretion  rate from
the inner  accrection disc, even when  the rate of  mass-injection into the
inner  part of  the disc  is random.   The predicted  X-ray  power spectrum
exhibits  a  power-law  index  of  about  1.6, with  a  flattening  at  low
frequencies.   This model  can possibly  be applied  to the  unified model,
thereby providing the  mechanism to produce changes in  the accretion rate.
Whether or not  it can produce a large enough  range in mass-accretion rate
(0.7--1.02 times the Eddington limit;  Psaltis et~al. \cy{plm95}) and a low
enough break frequency (Sco~X-1 is  predicted to have a turnover time-scale
of several days; Dieters \& van~der~Klis \cy{dv00}) needs to be tested.

An  alternative, yet  similar, model  for VLFN  is the  `dripping handrail'
model (e.g.,  Young \&  Scargle \cy{ys96}), in  which the amount  of matter
present at  the inner  edge of  the accretion disc  varies both  quasi- and
aperiodically.  If luminosity is considered  to be proportional to the mass
at the inner edge of the disc,  then both a power-law variability and a QPO
are produced.  The frequency of  the QPO is proportional to accretion rate,
so it  cannot be identified  as flaring branch  noise (see also  Dieters \&
van~der~Klis \cy{dv00}).   If the rate at which  matter \emph{leaves\/} the
disc   is    considered,   then   the    power-law   component   disappears
(Steiman-Cameron  et~al.   \cy{sys94}).    This  model  cannot,  therefore,
produce a power-law variability in mass-accretion rate, and so it cannot be
applied to the unified model.

Time dependent nuclear burning on the  surface of the neutron star has also
been  proposed  to produce  VLFN,  with  the  power-law index  expected  to
increase with increasing  mass-accretion rate (Bildsten \cy{b95}).  Dieters
\&  van~der~Klis (\cy{dv00}) noted  that: the  predicted variability  has a
characteristic  time-scale, in contrast  to the  observed VLFN  in Sco~X-1,
which is a power-law over time-scales  ranging from 10~s to about 14~h; and
the energy available from nuclear  burning is not sufficient to produce the
observed fractional rms amplitude.   Therefore, it is difficult for nuclear
burning alone to account for VLFN.

Our  results provide  further  evidence that  the  variations in  intensity
comprising the very  low frequency noise are associated  with movement that
is not directly along the Z track.  A possible scenario is that changes in
mass-accretion  rate,  or  $\eta$,  produce  motion in  which  there  is  a
component  both along,  and across,  the  track, and  that the  cross-track
component is more prominent  at higher frequencies.  Alternatively, changes
in  mass-accretion rate  or $\eta$  might only  produce movement  along the
track, with nuclear burning  providing a cross-track component.  Except for
the transition from  the NB into the FB, we observed  a general decrease in
power-law index with increasing inferred mass-accretion rate.  In contrast,
the nuclear  burning model  predicts an increase  in index  with increasing
mass-accretion  rate.   Therefore,  if  nuclear  burning  is  found  to  be
contributing significantly to  VLFN, the `$\eta$ model' for  the Z-track is
preferred, since it  does not specify a particular  mass-accretion rate for
each  part  of  the track.   As  pointed  out  by Dieters  \&  van~der~Klis
(\cy{dv00}), the two types of  variations would need to combine and produce
the observed VLFN power-law.

\section{Conclusions}

We have  carried out  the most  comprehensive study to  date, of  the X-ray
fast-time variability of Sco~X-2 as a  function of position on the Z track.
We found low frequency noise in  the NB, and typical Sco~X-2 flaring branch
noise in  the FB,  with centroid frequencies  in the range  3.3--5.8~Hz and
5.4--7.6~Hz, respectively.   The LFN  was strongest at  intermediate photon
energies,  while FBN was  hard. The  various models  seeking to  provide an
explanantion for LFN in Z sources must account for the energy dependence we
observed.   With regard  to  FBN,  we confirm  the  conclusions reached  by
O'Neill et~al.  (\cy{oks01}), and again note the similarity between FBN and
the  peaked  noise  seen the  FB  of  Cyg~X-2  at low  overall  intensities
(Kuulkers et~al.  \cy{kwv99}).  We suggest that an adaptation of the photon
bubble  model   may  possibly  provide   an  explanation  for   FBN  (Klein
et~al. \cy{kaj96a}).

We found  a new  peaked noise  feature in the  NB and  FB, with  a centroid
frequency in the range 11--54~Hz.  This high frequency peak was stronger at
higher  energies.   It  could   be  fitted  satisfactorily  with  a  single
Lorentzian, but,  in the NB,  it could also  be fitted with  two Lorentzian
components, representing a possible HBO and sub-HBO peak.

We found  very low  frequency noise throughout  the Z track.   It exhibited
abrupt changes  in the  power-law index and  fractional rms  amplitude with
movement  from the NB  into the  FB.  We  calculated complex  cross spectra
between intensity  and hardness, and  discovered that VLFN is  not entirely
due to movement along the Z track.  We speculate that VLFN may be due to:
variations in mass-accretion rate or  $\eta$ that produce motion in the HID
with  a component  both parallel,  and perpendicular,  to the  track;  or a
combination  of nuclear  burning on  the  neutron star,  and variations  in
accretion or $\eta$.

\section*{Acknowledgments}

This  research  has  made  use  of  data  obtained  from  the  High  Energy
Astrophysics  Science Archive  Research Center,  NASA/Goddard  Space Flight
Center.  The authors thank Alan Smale for assistance in obtaining the data.
PMO has in part been supported by the Australian Postgraduate Award scheme. 
This  work  was supported  in  part  by  the Netherlands  Organization  for
Scientific Research (NWO).

\appendix

\section{Correcting \textbfit{RXTE\/} data for differential deadtime
and channel cross-talk \label{sect:app}}

In the case of {\it RXTE\/}, the observed counting rate \emph{per detector}
$\mu$,  and incident  counting rate  $\lambda$,  can be  divided into  four
components\footnote{http://rxte.gsfc.nasa.gov/docs/xte/recipes/pca\_deadtime.html}:`Good
Xenon'   counting   rate   $\mu_{\mathrm{xe}}$;   the   `Remaining   Count'
$\mu_{\mathrm{r}}$;   the   counting    rate   from   the   propane   layer
$\mu_{\mathrm{p}}$;   and   the    `Very   Large   Event'   counting   rate
$\mu_{\mathrm{v}}$.

The counting rate of Good Xenon  events is the X-ray counting rate from the
source and background.  The Remaining  Count is the counting rate of events
that are detected simultaneously at  more than one anode.  These coincident
events are primarily due to particles, and, in bright sources, may also due
to `genuine' photons.  In each detector,  there is a propane layer in front
of the xenon  layers which is used primarily  for anti-coincidence, and may
also  be  used  to detect  source  photons  in  the range  1--3~keV  (Bradt
et~al. \cy{brs93}).  The Very Large Event counting rate is from events that
are  above  the  upper  discriminator;  these  events  are  mainly  due  to
particles.

Each  component contributes  to the  detector deadtime.   The  deadtime per
event  from   each  component  is:  $\tau_{\mathrm{xe}}   \sim  10~\mu  s$,
$\tau_{\mathrm{r}} \sim  10~\mu s$, $\tau_{\mathrm{p}} \sim  10~\mu s$, and
$\tau_{\mathrm{v}}  =  150~\mu  s$  ($\tau_{\mathrm{v}}$ is  a  commandable
instrument parameter).

The incident and observed counting rates are related by

\begin{equation}
\mu = f \lambda \label{eqn:a}
\end{equation}

\noindent where $f$ is the deadtime correction factor, and is given by

\begin{equation}
f       =       1       -       \tau_{\mathrm{xe}}\mu_{\mathrm{xe}}       -
\tau_{\mathrm{r}}\mu_{\mathrm{r}}   -  \tau_{\mathrm{p}}\mu_{\mathrm{p}}  -
\tau_{\mathrm{v}}\mu_{\mathrm{v}}
\end{equation}

\noindent and

\begin{equation}
f     =    \frac{1}{1    +     \tau_{\mathrm{xe}}\lambda_{\mathrm{xe}}    +
\tau_{\mathrm{r}}\lambda_{\mathrm{r}}+\tau_{\mathrm{p}}\lambda_{\mathrm{p}}
+ \tau_{\mathrm{v}}\lambda_{\mathrm{v}}}
\end{equation}

The  actual fractional  rms  amplitude $r_{\mathrm{xe}}$  of variations  in
$\lambda_{\mathrm{xe}}$  may  be related  to  the  observed fractional  rms
amplitude  $r'_{\mathrm{xe}}$ of variations  in $\mu_{\mathrm{xe}}$  by the
following expression (van~der~Klis \cy{v89})

\begin{equation}
r_{\mathrm{xe}} = \frac{r'_{\mathrm{xe}}}{C}
\end{equation}

\noindent where

\begin{equation}
C       =       \left|       1      +       \frac{\lambda_{\mathrm{xe}}}{f}
\frac{df}{d\lambda_{\mathrm{xe}}} \right|
\end{equation}

\begin{figure}
\begin{center}
\epsfig{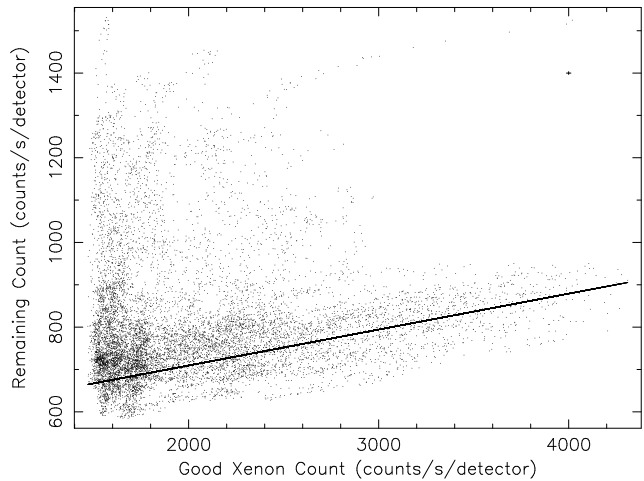}
\caption{Deadtime corrected Remaining Count versus deadtime corrected
  Good  Xenon counting  rate.  A  representative error  bar is  shown in
  the upper right.  The solid line shows  the best-fitting model fitted in
  the Good Xenon counting range 3000--4600~counts~s$^{-1}$~detector$^{-1}$,
  and extrapolated to lower counting rates.}
\label{fig:RemainingCnt}
\end{center}
\end{figure}

\begin{figure}
\begin{center}
\epsfig{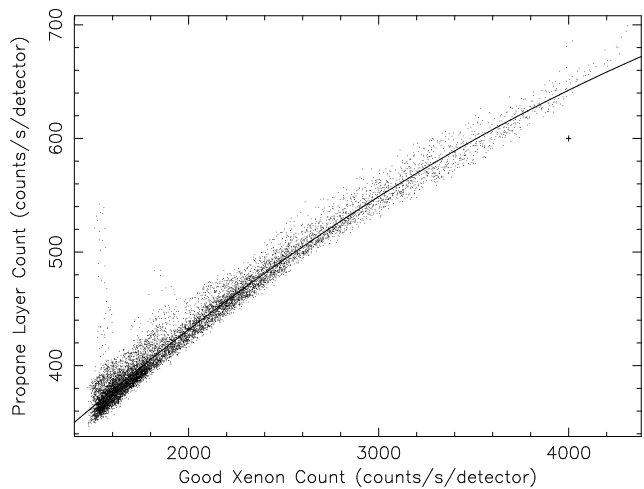}
\caption{Deadtime corrected propane layer counting rate versus deadtime
  corrected Good Xenon counting rate.  A representative error bar is shown
  in the upper right. The solid line shows the best-fitting model.}
\label{fig:VpCnt}
\end{center}
\end{figure}

\begin{figure}
\begin{center}
\epsfig{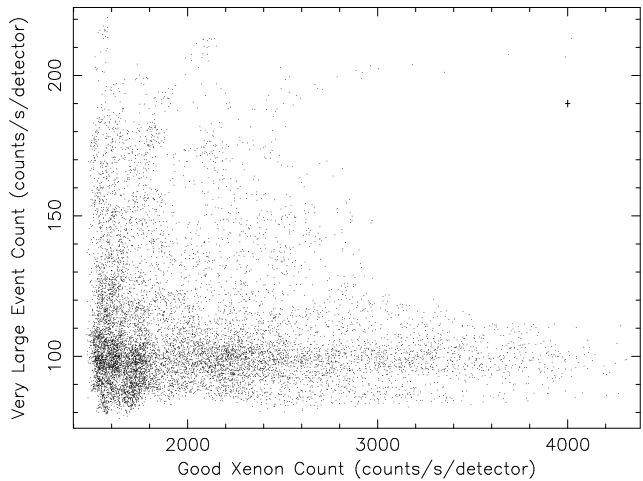}
\caption{Deadtime corrected Very Large Event counting rate versus deadtime
  corrected  Good  Xenon counting  rate.   A  representative  error bar  is
shown in the upper right.}
\label{fig:VLECnt}
\end{center}
\end{figure}

To   determine  $C$,   we   need  to   evaluate   $df  /   d
\lambda_{\mathrm{xe}}$.    Therefore,  we  also   need  to   determine  how
$\lambda_{\mathrm{r}}$,  $\lambda_{\mathrm{p}}$ and $\lambda_{\mathrm{v}}$,
depend   on   $\lambda_{\mathrm{xe}}$.   In   Figs.~\ref{fig:RemainingCnt},
\ref{fig:VpCnt},    and    \ref{fig:VLECnt},    we   present    plots    of
$\lambda_{\mathrm{r}}$, $\lambda_{\mathrm{p}}$, and $\lambda_{\mathrm{v}}$,
respectively, versus $\lambda_{\mathrm{xe}}$.   The data presented are from
Standard~1 mode using 16~s averages.

In~Fig.~\ref{fig:RemainingCnt},        in         the        region
$\lambda_{\mathrm{xe}}>$3000~counts~s$^{-1}$~detector$^{-1}$,  there  is  a
roughly  linear  relationship between  the  Good  Xenon  counting rate  and
Remaining            Count.             In            the            region
$\lambda_{\mathrm{xe}}<$3000~counts~s$^{-1}$~detector$^{-1}$,   the   lower
envelope of  the plot also  follows the linear relationship  observed above
3000~counts~s$^{-1}$~detector$^{-1}$,  suggesting   that  the  relationship
holds   for   all   $\lambda_{\mathrm{xe}}$.    The   best-fitting   linear
coefficient,    as   fitted    in    the   $\lambda_{\mathrm{xe}}$    range
3000--4600~counts~s$^{-1}$~detector$^{-1}$,  was 0.085.  The  large scatter
of   points  lying   above   the  linear   relationship,   in  the   region
$\lambda_{\mathrm{xe}}<$3000~counts~s$^{-1}$~detector$^{-1}$,     can    be
attributed to  events from particles.  This interpretation  is supported by
the presence of a similar scattering of points in Fig~\ref{fig:VLECnt}.

In  Fig.~\ref{fig:VpCnt},   there  is  a   quadratic  relationship  between
$\lambda_{\mathrm{p}}$   and  $\lambda_{\mathrm{xe}}$.    The  best-fitting
quadratic and linear coefficients were $-$1.165e-5 and 0.175, respectively.
In   Fig.~\ref{fig:VLECnt},   there  is   no   clear  correlation   between
$\lambda_{\mathrm{v}}$ and $\lambda_{\mathrm{xe}}$.

From the plots  presented in Figs.~\ref{fig:RemainingCnt}, \ref{fig:VpCnt},
and \ref{fig:VLECnt}, we find

\begin{align}
\frac{d  \lambda_{\mathrm{r}}}{d \lambda_{\mathrm{xe}}}  &=
0.085 \\
\frac{d  \lambda_{\mathrm{p}}}{d \lambda_{\mathrm{xe}}}  &=
-(2.33\mathrm{e-}5)\lambda_{\mathrm{xe}} + 0.175  \\
\frac{d  \lambda_{\mathrm{v}}}{d \lambda_{\mathrm{xe}}}  &=
0
\end{align}

Now we can evaluate $d f / d \lambda_{\mathrm{xe}}$ and find $C$

\begin{align}
\frac{d f}{d \lambda_{\mathrm{xe}}} &= \frac{
-\tau_{\mathrm{xe}}\left[1      +      0.085      -      (2.33\mathrm{e-}5)
  \lambda_{\mathrm{xe}} + 0.175 \right] }{\left( 1 +
\tau_{\mathrm{xe}}\lambda_{\mathrm{xe}}                                    +
\tau_{\mathrm{r}}\lambda_{\mathrm{r}}
+\tau_{\mathrm{p}}\lambda_{\mathrm{p}}                                    +
\tau_{\mathrm{v}}\lambda_{\mathrm{v}}            \right)^{2}}            \\
\frac{d f}{d        \lambda_{\mathrm{xe}}}        &=        -
\tau_{\mathrm{xe}}[1.26 - (2.33\mathrm{e-}5)\lambda_{\mathrm{xe}}] {f^{2}}
\end{align}

\noindent And thus

\begin{equation}
C = \left| 1 - \tau_{\mathrm{xe}}[1.26 -
  (2.33\mathrm{e-}5)\lambda_{\mathrm{xe}}] \lambda_{\mathrm{xe}} f
\right| 
\end{equation}

We used the above expression to correct  for differential deadtime and
channel cross-talk in the standard manner (e.g., Lewin et~al. \cy{llt92}).

\bsp

\label{lastpage}

\end{document}